\documentclass[11pt]{article}
\pdfoutput=1
\usepackage{jheppub}
\usepackage{float, extarrows, tikz-cd}
\usepackage{graphicx}
\usepackage{slashed}
\usepackage{tabularx,ragged2e}
\usepackage{amssymb}
\usepackage{subcaption}
\usepackage{amsmath,amssymb}
\usepackage{slashed}
\usepackage{caption}
\usepackage{xcolor}
\usepackage{dsfont}
\usepackage{verbatim}
\usepackage{mathtools, xcolor,ytableau, amsfonts,tikz}
\usepackage{graphicx}
\usepackage{physics}
\usepackage{simpler-wick}
\newcolumntype{C}{>{\Centering\arraybackslash}X}

\numberwithin{equation}{section}

\newcommand{\mO}{\mathcal{O}}

\newcommand{\pd}{\partial}

\def\le{\left}
\def\ri{\right}

\newcommand\p{\ensuremath{\partial}}
\newcommand{\es}[2] {\begin{equation} \label{#1} \begin{split} #2 \end{split} \end{equation}}

\def\<{\langle}
\def\>{\rangle}

\newcommand\ep{\epsilon}
\newcommand\sig{\sigma}
\newcommand\lam{\lambda}
\newcommand\Lam{\Lambda}
\newcommand\om{\omega}

\newcommand\ga{{\ensuremath{{\gamma}}}}

\newcommand\de{{\ensuremath{{\delta}}}}

\title{Boundary Conformal Field Theory at Large Charge} 
\author[a,b]{Gabriel Cuomo,}   
     \author[a]{M\'ark Mezei,}                                       
              \author[a]{Avia Raviv-Moshe}   
              \affiliation[a]{Simons Center for Geometry and Physics, SUNY, Stony Brook, NY 11794, USA}      
              \affiliation[b]{C. N. Yang Institute for Theoretical Physics, Stony Brook University, Stony Brook, NY 11794, USA}                                                              
                                            
\emailAdd{gcuomo@scgp.stonybrook.edu}					 
     \emailAdd{mmezei@scgp.stonybrook.edu}   
          \emailAdd{araviv-moshe@scgp.stonybrook.edu}
                                           
\abstract{We study operators with large internal charge in boundary conformal field theories (BCFTs) with internal symmetries. Using the state-operator correspondence and the existence of a macroscopic limit, we find a non-trivial relation between the scaling dimension of the lowest dimensional CFT and BCFT charged operators to leading order in the charge.  We also construct the superfluid effective field theory for theories with boundaries and use it to systematically calculate the BCFT spectrum in a systematic expansion. We verify explicitly many of the predictions from the EFT analysis in concrete examples including the classical conformal scalar field with a $|\phi|^6$ interaction in three dimensions and the $O(2)$ Wilson-Fisher model near four dimensions in the presence of boundaries.  In the appendices we additionally discuss a systematic background field approach towards Ward identities in general boundary and defect conformal field theories,  and clarify its relation with Noether's theorem in perturbative theories.}

\begin{document} 
\maketitle

\section{Introduction and summary}

Various quantum field theories that are strongly coupled  sometimes acquire  significant simplification under an expansion in some controlling parameter.  This is especially relevant in the context of conformal field theories (CFTs), where recent studies have shown the existence of such simplification in sectors that are characterized by large quantum numbers. Examples include CFTs in the regimes of large spin~\cite{Alday:2007mf,Fitzpatrick:2012yx,Komargodski:2012ek,Caron-Huot:2017vep}, large scaling dimensions~\cite{Lashkari:2016vgj,Cardy:2017qhl,Alday:2019qrf,Delacretaz:2020nit}, and large global charges~\cite{Hellerman:2015nra,MoninCFT,Jafferis:2017zna,Hellerman:2017veg}. In this paper we will focus on the third case, and consider conformal field theories with an additional global symmetry, focusing mostly on a global $U(1)$.  We will focus in particular on the implications of a boundary for the large charge sector of the theory.

Quantum field theories on manifolds with a boundary are known to have important applications ranging from condensed matter physics to cosmology and string theory. The study of conformal field theories in the presence of non-trivial boundaries has attracted much attention in recent years (see e.g.~
\cite{Ishibashi:1988kg,Cardy:1989ir,Affleck:1991tk,Friedan:2003yc,Cardy:2004hm,McAvity:1993ue,McAvity:1995zd,Takayanagi:2011zk,
Jensen:2015swa,
Casini:2016fgb,Herzog:2017xha,Andrei:2018die,DiPietro:2019hqe}
 and references therein).  In particular it has been shown that BCFTs can be systematically studied within a conformal bootstrap approach
 ~\cite{Liendo:2012hy,Gliozzi:2015qsa,Billo:2016cpy,Antunes:2021qpy},  similar to the one usually adopted in standard CFTs~\cite{Belavin:1984vu,Rattazzi:2008pe,Poland:2018epd}.

Starting from a local CFT in $d$ dimensions with an internal symmetry group $G$, we obtain a boundary conformal field theory (BCFT) by considering the CFT in the half-space $x^d> 0$ coupled to a plane boundary at $x^d=0$ in such a way that the $(d-1)$-dimensional conformal group $SO(d,1)$ is preserved. The boundary in general also preserves a (possibly trivial) subgroup of the internal symmetry $H\subset G$.\footnote{It is also possible to consider boundaries with additional internal symmetries, under which only boundary degrees of freedom are charged; we will comment on examples of this kind in section~\ref{SecOther}.} In this work we are interested in correlation functions of boundary operators with large quantum numbers under the unbroken group $H$. 

The BCFT spectrum is given by the set of scaling dimensions of boundary operators, and it can therefore be recovered simply from boundary correlators. From the bootstrap viewpoint, correlation functions at the boundary obey all the standard axioms but for the absence of a conserved boundary stress tensor~\cite{Liendo:2012hy}. This lack of locality might seem to provide an obstacle towards applying the ideas of~\cite{Hellerman:2015nra} in this setup. As in that work, the situation is illuminated by the state-operator correspondence.

Let us first review the basic observation underlying the analysis of~\cite{Hellerman:2015nra} for CFTs on the plane with no boundaries. 
Consider a $d$-dimensional CFT with $U(1)$ symmetry. By virtue of the state-operator correspondence, the operator $\mO_Q$ with lowest scaling dimension $\Delta_Q$ for a fixed $U(1)$-charge $Q$ generically corresponds to a state $\ket{Q}$ with homogeneous charge density in the theory compactified on the cylinder with radius $R$. For $Q\gg 1$ the charge density $\rho\sim Q/R^{d-1}$ introduces a dimensionful parameter $\mu\sim\rho^{1/(d-1)}$ much larger than the geometric scale $1/R$. The large separation of scales indicates that the state $\ket{Q}$ and its nearby excitation can be naturally associated with a \emph{condensed matter phase} of the theory~\cite{Nicolis:2015sra}. Dimensional analysis then generically\footnote{This relation is sometimes violated in theories which obey a different macroscopic limit; this normally happens in theories with flat directions, e.g. in free theories or in SCFTs~\cite{Hellerman:2017veg,Jafferis:2017zna}.} dictates the relation $\varepsilon\simeq \alpha\rho^{\frac{d}{d-1}}$ between the energy and the charge density of the state, which translates into the result $\Delta_Q\simeq \alpha Q^{\frac{d}{d-1}}$ for the scaling dimension of the operator $\mO_Q$ for $Q\gg 1$. Under the additional assumption that the theory is found in a \emph{superfluid phase}, corrections to the result for $\Delta_Q$ and other CFT data can be studied within the framework of \emph{effective field theory} (EFT). In the EFT, the derivative expansion coincides with an expansion in inverse powers of the charge~\cite{Hellerman:2015nra,MoninCFT}. The EFT predictions have been verified in several weakly coupled models (see e.g.~\cite{delaFuente:2018qwv,Alvarez-Gaume:2019biu,Badel,Antipin:2020abu}), and are compatible with the results of Monte-Carlo calculations of $\Delta_Q$ in the critical $O(2)$ and $O(4)$ models in three dimensions~\cite{Banerjee:2017fcx,Banerjee:2019jpw}.

A similar picture holds when considering the theory in the presence of a boundary. 
For the case of BCFTs,  Weyl invariance allows to map the theory from the half-plane  to the strip $\mathds{R}\times \mathds{H} S^{d-1}$ geometry,  where $ \mathds{H} S^{d-1}$ denotes the $(d-1)$-dimensional hemisphere, in such a way that the boundary is mapped to the $\theta=\frac{\pi}{2}$ equator in (hyper-)spherical coordinates.  As in the usual plane-cylinder map, dilations on the plane are mapped to time translations on the strip. Therefore the spectrum of boundary operator dimensions on the plane agrees with the energy spectrum for the theory quantized on $\mathds{R}\times \mathds{H} S^{d-1}$.  We will use this correspondence extensively throughout the remainder of this paper.
We expect that the lowest dimensional boundary operator $\hat{\mO}_Q$ with charge $Q\gg 1$ creates a state whose charge density is approximately homogeneous, at least at distances $\ell\gg 1/\mu$ away from the boundary, for the theory on the strip $\mathds{R}\times\mathds{H}S^{d-1}$.
 The energy density of the state obeys the same relation $\varepsilon\simeq \alpha\rho^{\frac{d}{d-1}}$ as in the case without a boundary. Integrating this relation on the hemisphere, it follows that the scaling dimension $\hat{\Delta}_Q$ of the boundary operator can be naturally related to that of a bulk operator with charge $2Q$:
\begin{equation}\label{eq_DeltaQ_intro}
\hat{\Delta}_Q\simeq \frac12 \Delta_{2Q}\simeq \frac{\alpha}{2}(2 Q)^{\frac{d}{d-1}}\quad\text{for}\quad Q\gg 1\,.
\end{equation}
Throughout this paper, we will verify the above relation in the large $Q$ regime in various examples. 
Similarly,  if the bulk operator $\mO_Q$ corresponds to a superfluid state, the large charge sector of the BCFT will also be in a superfluid phase, whose properties can be systematically studied within EFT.  

Note that the above discussion is independent of the nature of the boundary,  which affects only the subleading corrections to eq.~\eqref{eq_DeltaQ_intro}.
In this work we apply a systematic EFT approach to boundary conditions in the superfluid effective theory to parametrize these corrections. Perhaps surprisingly,  we find that to leading order these always reduce to Neumann conditions for the global symmetry current, with the first $\mu^{-1}$ correction controlled by a single Wilson coefficient.  Physically, this parameter controls the charge accumulation or decrease towards the boundary. We demonstrate our ideas in the classical three-dimensional $|\phi|^6$ model in sec.~\ref{SecInvitation}, where we construct the corresponding finite density EFT both for Neumann and Dirichlet boundary conditions on the fundamental field $\phi$. A similar EFT approach can be applied to other phases of matter as well (such as those classified~\cite{Nicolis:2015sra}), and we show this explicitly for isotropic solids in the vicinity of a wall in sec.~\ref{SecOther2}.

In sec.~\ref{Sec_EFT} we then use the EFT to calculate the BCFT spectrum of large charge operators,  including excited states, and relate the results with the predictions for bulk operators of~\cite{Hellerman:2015nra}.
We show that the leading order quantum corrections contribute a universal logarithmic correction to the scaling dimensions of the lowest dimension boundary operators, and calculate its coefficient for various numbers of spacetime dimensions. 
In sec.~\ref{sec:O2BCFTinEpsilonExpansion} we demonstrate the validity of the predictions made from the superfluid boundary EFT analysis in the concrete example of the $O(2)$ model in $4-\varepsilon$ dimensions with both Neumann and Dirichlet boundary conditions, working in the large charge double-scaling limit of~\cite{Badel}.  In particular, comparing with the previous results of~\cite{Badel}, we find that the relation \eqref{eq_DeltaQ_intro}, and its subleading corrections as discussed in sec.~\ref{Sec_EFT}, are perfectly reproduced in the limit $\varepsilon Q\gg 1$ (with all Wilson coefficients determined). In the opposite limit $\varepsilon Q\ll 1$ our results instead perfectly agree with the outcome of diagrammatic calculations.

The paper is organized as follows. In section \ref{SecInvitationMain} after a lightning review of BCFT, we provide some physical intuition  by studying the conformal complex scalar field model with a sextic interaction in three dimensions in the half-plane. In section \ref{Sec_EFT} we construct the superfluid effective field theory in the presence of a boundary and calculate the BCFT spectrum of large charge operators.  In section \ref{sec:O2BCFTinEpsilonExpansion} we study the weakly coupled example of the $O(2)$ BCFT in $4-\varepsilon$ dimensions, both with Neumann and Dirichlet boundary conditions. In section \ref{SecOther} we discuss some other large charge phases of BCFTs, including the free charged scalar with an interacting boundary, free fermions, and theories with charged degrees of freedom only at the boundary. In section \ref{SecOther2} we comment on some additional applications including defect CFTs, the thermodynamic limit in BCFTs and boundary conditions in the solid EFT.  Some technical details can be found in appendices \ref{AppBdryAction} and \ref{App_D}. Appendix \ref{AppendixWardIdentity} and \ref{App_Dual_Bactions} instead contain general considerations regarding BCFTs and may be read independently from the rest of the text. In particular appendix \ref{AppendixWardIdentity} describes a systematic background field approach towards Ward-identities in boundary and defect CFTs. There we also clarify how the absence of a boundary stress tensor or current is compatible with perturbation theory in theories with non-trivial boundary conditions in weakly coupled BCFTs. In appendix \ref{App_Dual_Bactions} we comment on the possibility of having dual boundary actions for mixed boundary conditions in scalar quantum field theories.

\section{Invitation}\label{SecInvitationMain}

\subsection{Lightning BCFT review}

For future reference throughout the paper, we briefly review the basic properties of correlation functions of local operators in BCFTs following~\cite{Gliozzi:2015qsa}.
Correlation functions of local boundary operators are completely specified by the spectrum of boundary scale dimensions $\{\hat{\Delta}_i\}$ and three-point coefficients $\{\hat{\lambda}_{ijk}\}$, which specify the boundary OPE as in usual CFTs.  Bulk operators, besides obeying the usual bulk OPE,  can be decomposed into boundary operators using the bulk-to-boundary OPE~\cite{McAvity:1995zd}
\begin{equation}\label{eq_intro_bulk_to_bdry}
\mO_i(x^a,x^d)=\frac{a_{i}}{|2 x^d|^{\Delta_i}}+\sum_{k}
\frac{b_{i k}}{|2x^d|^{\Delta_{i}-\hat{\Delta}_k}}
D\left[x^d,\pd_b\right]\hat{\mO}_k(x^a)\,,
\end{equation}
where we denoted boundary operators with a hat and $x^a$ denotes coordinates parallel to the plane $a=1,\ldots,{d-1}$.  Notice that the conformal symmetry implies that only scalar operators may have a non-zero one-point function $a_{i}\neq 0$.  We therefore conclude that the set $\{\Delta_i,\hat{\Delta}_i,a_i,b_{ik},\hat{\lambda}_{ijk}\}$ completely specifies the local observables of the theory.\footnote{More precisely this is true up to a finite set of \emph{central charges} associated with operators whose normalization is fixed by Ward identities, such as the stress tensor, internal currents and the displacement operator discussed below.} Notice that this set does not include the bulk OPE coefficients.

The constraints of locality and conformal invariance on BCFTs were studied by several  authors (see e.g. \cite{McAvity:1993ue,Jensen:2015swa,Billo:2016cpy}). In appendix \ref{AppendixWardIdentity} we provide a comprehensive review of the relevant Ward identities and their derivation. Here we highlight some relevant results.

Generically, BCFTs do not admit a conserved boundary stress tensor. Indeed, this would imply the existence of additional spacetime conserved charges besides those constructed from the bulk stress tensor.  Similarly, the BCFT does not have any conserved boundary current in the absence of internal symmetries under which only boundary degrees of freedom are charged.  In turn the breaking of translations in the $d$th direction implies the existence of a boundary scalar operator $\hat{D}$ of dimension $d$. This is the unique scalar in the bulk to boundary OPE of the stress tensor~\cite{McAvity:1993ue,DiPietro:2019hqe}:
\begin{equation}
T_{\mu\nu}(x^a,x^d)\stackrel{x^d\rightarrow 0}{\sim}\frac{d}{d-1}\left(
\delta_{\mu d}\delta_{\nu d}-\frac{\delta_{\mu\nu}}{d}\right)\hat{D}(x^a)
+\ldots\,.
\end{equation}
Similarly, when the boundary breaks an internal symmetry $G$ to a subgroup $H$, the BCFT contains $\text{rank}(G)-\text{rank}(H)$ scalar boundary operators with dimension $d-1$. These are the unique scalars in the bulk to boundary OPE of the broken currents~\cite{Herzog:2017xha}.

 \subsection{Superfluid on the half-plane}\label{SecInvitation}

Before analyzing the EFT describing BCFT states on the strip, here we would like to provide some physical insights on the role of boundary conditions in the superfluid theory.  To this aim, here we analyze a simple scale-invariant classical model consisting of a single complex field in the three-dimensional half plane
\begin{equation}\label{eq_ex}
S=\int_{y\geq 0} dt d^2x\left(|\pd\phi|^2-\frac{g^2}{6}|\phi|^6\right)\,.
\end{equation}
We work in Lorentzian signature and the integration is restricted to $y\geq 0$ in coordinates $x^\mu=(t,x,y)$.  For the sake of simplicity, we consider boundary conditions at $y=0$ linear in the field. There are therefore two options of preserving scale (and $U(1)$) symmetry, namely
\begin{equation}
\text{Neumann: }\pd_y\phi\vert_{y=0}=0\qquad
\text{or}\qquad
\text{Dirichlet: }\phi\vert_{y=0}=0\,.
\end{equation}

We want to consider the theory at finite charge density. By ensemble equivalence at infinite volume, this just amounts at turning on a non-zero chemical potential $\mu$. In this regime,  we expect the internal $U(1)$ symmetry to be spontaneously broken \cite{Son:2002zn} (together with other spacetime symmetries \cite{Nicolis:2015sra}); therefore, the low the low energy description on scales much larger than $1/\mu$ consists of a single shift invariant Goldstone boson.  In this section we study the effect of boundary conditions on this low energy EFT, by classically integrating out the radial mode.

It is convenient to decompose the field as $\phi=\frac{\rho}{\sqrt{2}}e^{i\pi}$. The action \eqref{eq_ex} in the presence of a chemical potential then reads:
\es{UVaction}{
S=\int_{y\geq 0} dt d^2x\left[\frac12(\pd\rho)^2+\frac12(\pd\chi)^2\rho^2-\frac{g^2}{48}\rho^6\right]\,,\qquad
\chi(x)=\mu t+\pi(x)\,.
}

We now analyze Neumann and Dirichlet boundary conditions separately. Consider first the simpler case of Neumann boundary conditions.  These can be written as:
\begin{equation}\label{eq_ex_Neumann}
\pd_y\rho\vert_{y=0}=j_y\vert_{y=0}=0\,,
\end{equation}
where $j_\mu=\pd_\mu\chi\,\rho^2$ is the $U(1)$ current. In this case we can safely integrate out the field $\rho$ using the equations of motion everywhere. Just as in the absence of a boundary, to leading order in derivatives this gives $\rho\simeq\frac{2^{3/4}\sqrt{\mu}}{\sqrt{g}}\simeq \frac{2^{3/4}}{\sqrt{g}}(\pd\chi)^{1/2}$ (with the shorthand notation $(\pd\chi)\equiv\sqrt{g^{\mu\nu}\p_\mu \chi \p_\nu \chi}$ ) and the low energy EFT reads:
\begin{equation}\label{eq_ex_Seft_Neu}
S_{EFT}^{(Neu)}=\frac{2\sqrt{2}}{3 g}\int_{y\geq 0} dt d^2x (\pd\chi)^3+\ldots\,.
\end{equation}
From the variation of $S_{EFT}$ we find the bulk equations of motions as well as the boundary conditions for the field $\chi$:
\begin{equation}
\pd_\mu j^\mu=0\quad\text{with}\quad
j_y\vert_{y=0}=0\,,
\end{equation}
where $j_\mu=\frac{2\sqrt{2}}{g}\pd_\mu\chi(\pd\chi)$ in the EFT.  Notice that the boundary condition $j_y\vert_{y=0}=0$ in the EFT precisely matches that of the UV theory \eqref{eq_ex_Neumann}.

\begin{figure}[t]
   \centering		
   \includegraphics[width=0.5\textwidth]{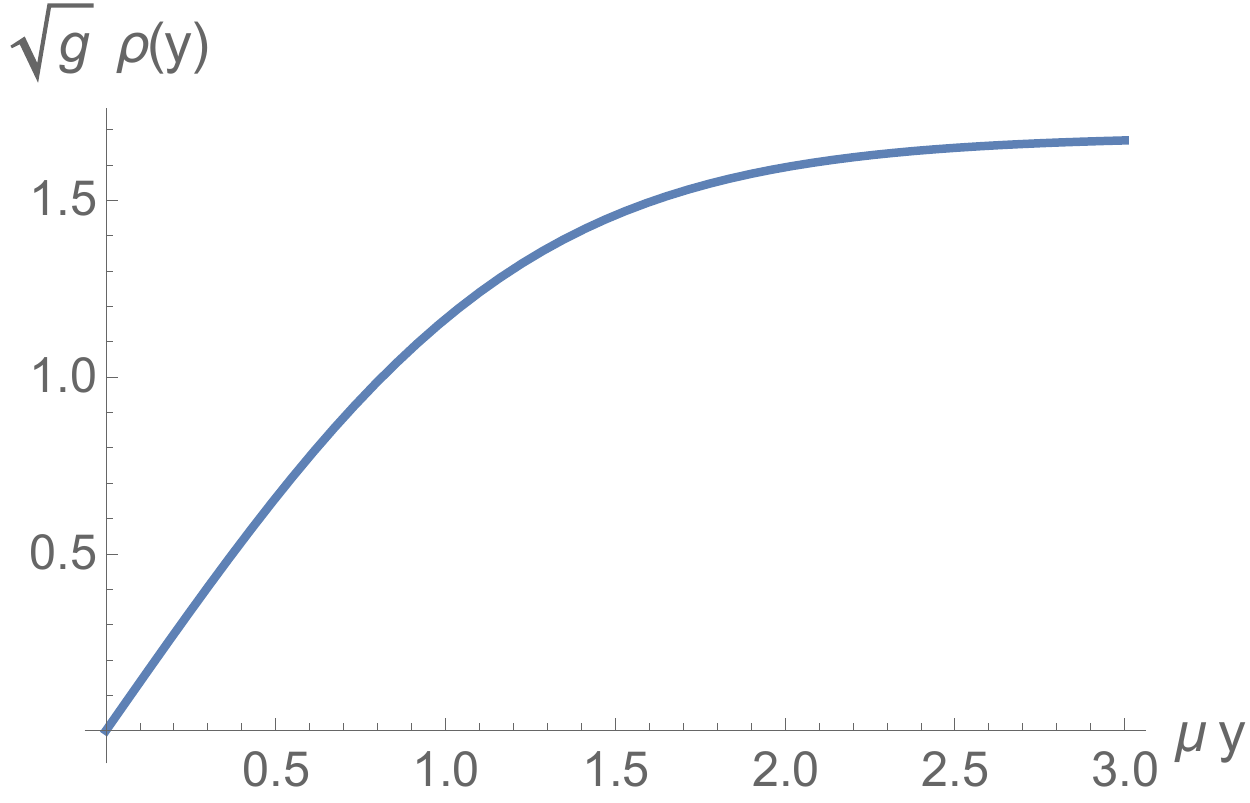}
        \caption{ The profile of the radial field $\rho(y)$ given in \eqref{eq_ex_rho} that satisfies Dirichlet boundary conditions.}    
\label{fig:profile}
\end{figure}

Consider now Dirichlet boundary conditions. In polar field coordinates these read:
\begin{equation}\label{Eq_ex_BC_Dir}
\rho\vert_{y=0}=0\,\quad\implies\quad j_{\mu}\vert_{y=0}=0\,,
\end{equation}
implying that the current vanishes at the boundary.  In this case a constant $\rho$ is incompatible with the boundary conditions, 
rather the classical background is obtained by solving the equations of motion in terms of a profile $\rho=\rho(y)$ which vanishes at $y=0$ and goes to a constant far from the boundary. The explicit solution reads:
\begin{equation}\label{eq_ex_rho}
\rho(y)=
\frac{2^{5/4}\sqrt{\mu}}{\sqrt{g}}\frac{  \tanh (\mu  y)}{\sqrt{3-\tanh ^2(\mu  y)}}
=\frac{2^{3/4}\sqrt{\mu}}{\sqrt{g}}\times
\begin{cases}\displaystyle
\sqrt{\frac23}\, \mu y-\frac{(\mu y)^3}{3\sqrt{6}}+\mO\left((\mu y)^5\right) 
& \text{for }y\ll\mu^{-1}\,,\\[5pt]
\displaystyle
1- 3e^{-2\mu y}+\mO\left(e^{-4 \mu y }\right) & \text{for }y\gg\mu^{-1}\,.
\end{cases}
\end{equation}
See fig.~\ref{fig:profile} for a plot of the profile. Note that the dimensionless product $\mu y$ controls the behavior of the solution. Sufficiently far from the boundary, $y\gg 1/\mu$, $\rho$ takes the same constant value as for the case of Neumann boundary conditions up to exponentially small corrections; therefore the bulk action of the low energy EFT is still given by \eqref{eq_ex_Seft_Neu}, independently of the boundary conditions, as expected from the locality of the theory. 

How about the boundary condition on the Goldstone field $\pi(x)$? Clearly we cannot impose the vanishing of all the components of the current as in the UV description \eqref{Eq_ex_BC_Dir}, since this would not allow for any non-trivial profile for $\pi(x)$. Physically, the charge density $j_0\simeq \mu \rho^2(y)$ decays non-trivially only for $y\ll 1/\mu$, while in the EFT we resolve the location of the boundary only up to an $\mO\left(1/\mu\right)$ uncertainty; therefore we cannot match its value in the EFT very close to the boundary.  Instead, rather than requiring that the local value of $j_0$ is the same in the UV description and in the EFT at the boundary, we should only match the surface charge density in the $x$ direction in a finite patch of length $L\gg 1/\mu$:
\begin{equation}\label{eq_ex_matching}
\int _0^L dy \,j_0^{(EFT)}(y)\stackrel{!}{=}
\int_0^L dy \,j_0^{(UV)}(y)=\frac{2\sqrt{2}}{g}\left[\mu ^2 L
+\frac{\sqrt{3}}{2}  \mu  \log \left(2-\sqrt{3}\right)
+\mO\left(e^{-L\mu}\right)
\right]\,.
\end{equation}
The leading piece proportional to $\mu^2L$ is reproduced by the constant charge density in the bulk. The first correction $\sim \mu$ is instead independent of the integration length $L$ and can therefore be \emph{effectively} mimicked by a negative (notice $ \log \left(2-\sqrt{3}\right)\simeq -1.3$) charge accumulation at the boundary.  This is achieved by adding to the action \eqref{eq_ex_Seft_Neu} the simplest conceivable boundary term:
\es{eq_ex_S_Dir}{
S_{EFT}^{(Dir)}&=\frac{2\sqrt{2}}{3 g}\int_{y\geq 0} dt d^2x (\pd\chi)^3+
\frac{b}{2g}\int_{y=0} dt dx(\hat{\pd}\chi)^2+
\ldots\,,\\[6pt]
b&=\sqrt{6}\log\left(2-\sqrt{3}\right)\simeq -3.2\,,
}
where $\hat{\pd}_a$ denotes the derivative along the directions parallel to the boundary $x^a=(t,x)$. Computing the charge density via the Noether procedure one finds:
\begin{equation}
j_\mu^{(EFT)}=\frac{1}{g}\left[2\sqrt{2}(\pd\chi)\pd_\mu\chi+\delta(y)\,b\delta_\mu^{a}\hat{\pd}_a\chi\right]\,,
\end{equation}
whose integral matches precisely eq.~\eqref{eq_ex_matching} on the background. Notice that the exponentially small corrections in eq.~\eqref{eq_ex_matching} instead are not  reproduced in the EFT. Finally the boundary condition on the Goldstone field is obtained from the variation of the action \eqref{eq_ex_S_Dir}:
\begin{equation}\label{eq_ex_bc}
\left[2\sqrt{2}(\pd\chi)\pd_y\chi -b\hat{\pd}^2\chi\right]_{y=0}=0\quad
\implies\quad
\pd_y\pi\vert_{y=0}\simeq\left.   \frac{b}{2\sqrt{2}}\frac{\hat{\pd}^2\pi}{\mu}\right\vert_{y=0}= {\cal O}\le(\frac{k^2}{ \mu}\ri)\,,
\end{equation}
where we expanded to leading order in field fluctuations in the last equation.  This makes it manifest that the boundary term in eq.~\eqref{eq_ex_S_Dir} can be thought as an $\mO\left(\hat{\pd}/\mu\right)$ correction. Therefore to leading order in derivatives the EFT describing the model eq.~\eqref{eq_ex} in the presence of a chemical potential is the same for both Neumann and Dirichlet, and more generally it is independent of the UV boundary conditions. 

As a non-trivial consistency check we can compute the phase shift induced by the boundary on an incoming plane wave in the EFT and check that it correctly reproduces the UV theory result. Namely, in the EFT we work at the linearized level and consider a plane-wave solution of the form:
\es{scatteringSol}{
\pi(x)=e^{-i\omega_k t-i k y}+e^{-i\omega_k t+i k y}e^{i\delta}\,,
}
where we consider $k>0$ and the dispersion relation gives $\omega_k=\frac{k}{\sqrt{2}}$, corresponding to the conformal speed of sound $c_s=1/\sqrt{2}$. The boundary condition \eqref{eq_ex_bc} then determines the phase shift $\delta$ as:
\es{PhaseShift}{
\delta=\frac{b \,k }{2\sqrt{2}\mu}+\mO\left(\frac{k^2}{\mu^2}\right)\,.
}
Since $b<0$, this results correspond to a negative time delay $d\delta/d k=\Delta t/c_s <0$. Physically this is because close to the boundary the charge density decreases and the wave approaches the speed of light $c_s=1$. 

We now compute the phase shift in the UV theory. We linearize the equations of motion for the fields around their background values using the Ansatz inspired by \eqref{scatteringSol}: $\chi=\mu t +\pi_\om(y) e^{-i\om t}$ and $\rho=\rho_0(y)+\rho_1(y)e^{-i\om t}$ with $\rho_0(y)$ given in \eqref{eq_ex_rho}. We can express
\es{rho1}{
\rho_1(y)={i/ (2\mu \om)}\le[\rho_0 \le(\om^2+{d^2/ dy^2}\ri) \pi_\om+2\rho_0' \pi'_\om\ri]\,,
}
and eliminate it from the problem at the expense of making the equation for $\pi_\om(y)$ fourth order in derivatives. This equation cannot be solved in terms of special functions, but its solution can be approximated 
 using matched asymptotic series expansions, as we summarize below.
 
 In the far region, for $y \mu\gg 1$, we can solve the equation in an expansion in $e^{-2\mu y}$. To leading order, $\rho_0$ is a constant, and $ \pi_\om(y)$ is a pure exponential. Since we are solving a fourth order equation, we get four branches of solutions, two are plane waves with $k(\om)$ and $-k(\om)$ and two are exponentials with growth/decay constant $c(\om)$. To set up the scattering problem, we discard the exponentially growing piece, and write the solution in the far region:
 \es{farRegion}{
 \pi_\om(y)=&e^{-ik y}\le[1+a(k)\,e^{-2\mu y}+\dots  \ri]\\
 &+e^{i\de}e^{ik y}\le[1+a(-k)\,e^{-2\mu y}+\dots  \ri]\\
 &+A e^{-c y}\le[1+\dots  \ri]\,,
 }
 where the dispersion relation (expanded in $\om/\mu$) is $k(\om)=\sqrt{2}\om-\om^3/(8\sqrt{2}\mu^2)+\dots$ and $c(\om)=2\mu+\dots$, and in \eqref{farRegion} we wrote out the terms that we included in our computation explicitly. 
 
 Near the boundary, we introduce a rescaled coordinate $Y\equiv \mu y$ and expand the resulting equation in $\om^2/\mu^2$. We solved the problem to subleading order in this expansion, $\pi_\om(Y)=B+\om^2/\mu^2\, \tilde \pi(Y)+\dots$, where $\tilde \pi(Y)$ has a lengthy expression including polylogarithms. In the near boundary expansion the general solution has four undetermined coefficients $c_i$:
 \es{pit}{
  \tilde \pi(Y)=\frac{c_1}{Y}+c_2+c_3 Y+c_4 Y^2+O(Y^3)\,.
 }
 The Dirichlet boundary condition on the full complex field gives the following two conditions for its real and imaginary parts:
 \es{2bcs}{
 \rho_1\vert_{y=0}=\le(\rho_0\, \pi_\om\ri)\vert_{y=0}=0\,,
 } 
 which fixes $c_1=c_3=0$.
 
The next step involves matching the far and near region solutions in their overlapping region of validity. $ \tilde \pi(Y)$ can be expanded for large $Y$ and matched to \eqref{farRegion} in a double expansion in $\mu$ to $O(1/\mu^2)$ and in $Y$ to $O(e^{-2Y})$.\footnote{We have determined the solution in the far region \eqref{farRegion}  to $O(e^{-6\mu y})$ and verified that all those extra terms match to the solution in the near region that was completely fixed from the lower order computation.} The phase shift that we set out to compute gets determined to leading order as  $\de={b\om/ (2\mu)}$ as in \eqref{PhaseShift} with the value of $b$ agrees with  \eqref{eq_ex_S_Dir}. This provides a highly nontrivial consistency check on the EFT.

\section{Large charge boundary operators from the superfluid EFT}\label{Sec_EFT}

\subsection{The EFT with a boundary}

Consider a BCFT in $d>2$ dimensions with $U(1)$ internal symmetry group. As emphasized in the introduction,  the simplest possibility is that operators with internal charge $Q\gg 1$ correspond to a superfluid phase of the theory on the strip $\mathcal{M}=\mathds{R}\times \mathds{H}S^{d-1}$ with radius $R$. The corresponding low energy EFT is written in terms of a single real Goldstone boson $\chi(x)=\mu t+\pi(x)$, and a $U(1)$ shift invariant action. As illustrated in the previous section, the action can be written in terms of a bulk and a boundary contribution:
\begin{equation}\label{eq_action_LO}
S=S_{bulk}+S_{bdry}\,.
\end{equation}
Due to the locality of the theory,  the bulk action does not depend on the boundary condition and coincides with the one constructed in~\cite{Hellerman:2015nra,MoninCFT}. Its form is fixed by $U(1)$ and Weyl invariance and up to second order in derivatives it reads:
\es{}{
S_{bulk}&=\int_{\mathcal{M}} d^dx\sqrt{g}\left\{c_1(\pd\chi)^d+
c_2\left[\mathcal{R}(\pd\chi)^{d-2}+\ldots\right] \right. \\ 
&\left.+c_3(\pd\chi)^{d-4}\left[\pd^\mu\chi\pd^\nu\chi
\mathcal{R}_{\mu\nu}+\ldots\right]+
\mO\left(\mathcal{R}^2(\pd\chi)^{d-4}\right)
\right\}\,, \label{eq_action_LO_bulk}
}
where here and the rest of the paper we use the shorthand $\sqrt{g}\equiv \sqrt{\abs{{\rm det}(g_{\mu\nu})}}$ and the dots stand for terms which vanish on the classical profile $\chi=\mu t$.  We are also assuming that the bulk theory is parity invariant for simplicity; for discussion of the parity breaking theories see~\cite{Cuomo:2021qws}. The Wilson coefficients $c_i$'s are the same as in the CFT without a boundary and they are expected to be $\mO(1)$ for strongly coupled theories, but may take parametrically large values in weakly coupled models. As in the example of sec.~\ref{SecInvitation}, we parametrize the boundary conditions by introducing the most general boundary action compatible with the symmetries. In appendix \ref{AppBdryAction} we show that $S_{bdry}$ depends on a unique Wilson coefficient up to corrections which are second order in derivatives:
\begin{equation}\label{eq_action_bdry_LO}
S_{bdry}=\int_{\pd\mathcal{M}} d^{d-1}x\sqrt{\hat{g}}\left[b_1(\hat{\pd}\chi)^{d-1}+\mO\left(\mathcal{R}(\pd\chi)^{d-3}\right)\right]\,,
\end{equation}
where $\hat{g}_{ab}$ is the induced metric on the boundary and $(\hat{\pd}\chi)^2=\hat{g}^{ab}\pd_a\chi\pd_b\chi$.  As explained in sec.~\ref{SecInvitation} the boundary equations of motion which follow from eq.~\eqref{eq_action_LO} can be thought as perturbations of the Neumann condition $n^\mu j_\mu\vert_{\pd\mathcal{M}}=0$, where $n_\mu$ is the normal to $\pd\mathcal{M}$.  

\subsection{The BCFT spectrum at large charge}

We want to use the action \eqref{eq_action_LO} to compute the BCFT spectrum of large charge operators. To this aim, it is useful to recall first the results for the theory on the cylinder (no boundary)~\cite{Hellerman:2015nra,MoninCFT}.  In this case, computing the Noether current from the bulk action on the classical profile $\chi=\mu t$, one finds the following relation between the total charge $Q$ and the chemical potential:
\begin{equation}\label{RmuConversionBulk}
R\mu= \left(\frac{Q}{c_1 d\Omega_{d-1}}\right)^{\frac{1}{d-1}}+
\mO\left(Q^{-\frac{1}{d-1}}\right)\,,
\end{equation}
where $\Omega_{d-1}=\frac{2\pi^{d/2}}{\Gamma(d/2)}$ is the volume of the $(d-1)$-sphere.  We then compute the energy of the ground-state by integrating the expectation value of the energy-momentum tensor. Neglecting momentarily quantum corrections, we find:
\begin{equation}
\begin{gathered}
\Delta_Q\vert_{classical}=\alpha_1 Q^{\frac{d}{d-1}}+\alpha_2Q^{\frac{d-2}{d-1}}+\ldots\,,\\[5pt]
\alpha_1=\frac{c_1(d-1) \Omega_{d-1}}{(c_1 d \Omega_{d-1})^{\frac{d}{d-1}}}\,,
\quad
\alpha_2=\frac{c_2 (d-2) (d-1) \Omega_{d-1}}{(c_1 d \Omega_{d-1})^{\frac{d-2}{d-1}}}\,, \label{eq_bulk_Delta}
\end{gathered}
\end{equation}
where the expansion runs in even powers of $R\mu\sim Q^{\frac{1}{d-1}}$ due to parity invariance.  Quantum corrections provide a $\mO\left(Q^0\right)$ contribution to \eqref{eq_bulk_Delta}. To compute them, we notice that the \emph{phonon} fluctuation field $\pi(x)=\chi(x)-\mu t$ obeys the following linearized equation of motion to leading order in derivatives
\begin{equation}\label{eq_EFT_lin_eq_bulk}
\ddot{\pi}-\frac{1}{d-1}\nabla_{S^{d-1}}^2\pi=0\,.
\end{equation}
The wave-solutions $e^{-i\omega_{\ell}t/R}f(\Omega)$ to the bulk equations have frequency
\begin{equation}\label{eq_EFT_omega_ell}
\omega_{\ell}=\sqrt{\frac{\ell(\ell+d-2)}{d-1}}\,,\qquad
\ell=1,2,\ldots\,,
\end{equation}
and are given by the (hyper)spherical harmonics with angular momentum $\ell$.  Therefore, at a quantum level, the spectrum of charge $Q$ operators can be organized as a Fock space in terms of single particle states with angular momentum $\ell$ and energy $\Delta_Q+\omega_{\ell}$.\footnote{The zero mode relates states with different charge~\cite{MoninCFT}.} The one-loop quantum correction to eq.~\eqref{eq_bulk_Delta} is given by the Casimir energy of the phonon and reads:
\begin{equation}\label{eq_bulk_Casimir}
\delta\Delta_Q^{(1-loop)}=\frac{1}{2}\sum_{\ell}n_{\ell}\omega_{\ell}\,,
\end{equation}
where $n_{\ell}=\frac{(2\ell+d-2)\Gamma(\ell+d-2)}{\Gamma(d-1)\Gamma(\ell+1)}$ is the dimension of the spin $\ell$-representation of $SO(d)$.  Upon regularizing the sum in eq.~\eqref{eq_bulk_Casimir} compatibly with all the symmetries, e.g. in dimensional regularization\footnote{See also~\cite{Gretsch:2013ooa} for details on dimensional regularization in theories with spontaneously broken conformal symmetry.} as in~\cite{Cuomo:2020rgt}, we find that the result is qualitatively different between even and odd-spacetime dimensions. This is because,  as it can be seen from \eqref{eq_bulk_Delta}, the classical result contains a $Q^0$ contribution only for $d=$even. Thus for $d=$odd the Casimir energy cannot be renormalized by any local counterterm; accordingly,  it is finite (when properly regularized) and the EFT predicts a theory-independent calculable $\mO(Q^0)$ contribution to $\Delta_Q$~\cite{Hellerman:2015nra}. Instead for even $d$ the Casimir energy is divergent and the EFT predicts the existence of a $Q^0\log Q$ term with universal coefficient~\cite{Cuomo:2020rgt}.  Overall, eq.~\eqref{eq_bulk_Casimir} leads to
\begin{equation}
\delta\Delta_Q^{(1-loop)}=\begin{cases}
c_d &\text{for }d=\text{odd}\,,\\
\gamma_d\log Q &\text{for }d=\text{even}\,,
\end{cases}
\end{equation}
where
\begin{equation}\label{EFTbulkQuantum}
c_d=\begin{cases}
-0.0937\ldots &d=3\\
- 0.1079\ldots &d=5\,,
\end{cases}\qquad
\gamma_d=
\begin{cases}
-\frac{1}{48\sqrt{3} } &d=4\\
- \frac{1}{60\sqrt{5}} &d=6\,.
\end{cases}
\end{equation}

We may now proceed in the same way to extract the spectrum in the presence of a boundary. In this case from the action \eqref{eq_action_LO} we extract the following relation between the chemical potential and the total charge $Q$ of the ground-state: 
\es{RmuConversion}{
R\mu= \left(\frac{2Q}{c_1 d\Omega_{d-1}}\right)^{\frac{1}{d-1}}-\frac{2b_1\Omega_{d-2}}{c_1 d\Omega_{d-1}}+
\mO\left(Q^{-\frac{1}{d-1}}\right)\,.
}
Besides the factors of $2$ arising from the reduced volume, the main difference with eq.~\eqref{RmuConversionBulk} is the second term.
This arises from the boundary current in the EFT, and its sign depends on whether the charge density increases or decreases close to the boundary.  In theories in which the current vanishes at the boundary,\footnote{Technically this means that its bulk to boundary OPE \eqref{eq_intro_bulk_to_bdry} is not singular, i.e. it does not contain any operator of dimension $\hat{\Delta}<d-1$.}  we naturally expect that the charge density decreases close to the boundary, corresponding to $b_1<0$ as in sec.~\ref{SecInvitation}.  This is indeed be the case for the $O(2)$ model in the epsilon expansion with Dirichlet boundary conditions, as we will show in sec.~\ref{sec:O2BCFTinEpsilonExpansion}.  In general however the sign of $b_1$ cannot be inferred \emph{a priori}; for instance we will show in the next section that quantum effects generate a coefficient $b_1>0$ in the epsilon expansion for the $O(2)$ model with Neumann boundary conditions.\footnote{For an  example of a classical model in which both signs occurs depending on the parameters, consider the following model consisting of two complex fields $\phi$ and $\psi$, one of which only lives at the boundary:
\begin{equation}
S=\int_{z\geq 0} d^4x\left[|\pd\phi|^2-\frac{\lambda}{4}|\phi|^4\right]+
\int_{z=0} d^3x\left[|\pd\psi|^2-g_1|\phi|^2|\psi|^2
-\frac{g_2}{2}\left(\phi^{*\,2}\psi^2+c.c.\right)-\frac{\tilde{\lambda}^2}{6}|\psi|^6\right]\,.
\end{equation}
The boundary conditions for $\phi$ are perturbations of Neumann.
Coupling this theory to a chemical potential $\mu$, it is easy to show that the low energy EFT is a superfluid, and that the sign of the coefficient $b_1$ in eq.~\eqref{eq_action_bdry_LO} is the same as that of $g_1+|g_2|$. Physically, while the bulk charge density carried by $\phi$ decreases close to the boundary, the field $\psi$ stores some charge (precisely at $z=0$).  The amount of charge stored by $\psi$ depends on the parameters and determines the sign of $b_1$ in the EFT.\label{fn:lag}} We will see in a moment that $b_1$ controls the difference $\hat{\Delta}_Q-\frac{1}{2}\Delta_{2Q}$.

As before, we compute the scaling dimension of the ground-state integrating the expectation value of the energy-momentum tensor. Neglecting momentarily quantum corrections, we find:
\begin{equation}\label{eq_deltabdry}
\hat{\Delta}_{Q}\vert_{classical}=\hat{\alpha}_1\left(2 Q\right)^{\frac{d}{d-1}}+
\hat{\beta}_1 (2 Q)+\hat{\alpha}_2\left(2 Q\right)^{\frac{d-2}{d-1}}+\ldots\,,
\end{equation}
where the coefficients $\hat{\alpha}_i$'s and $\hat{\beta}_i$'s depend on the Wilson coefficients in the action and can be related to those appearing in eq.~\eqref{eq_bulk_Delta} as:
\begin{equation}\label{eq_EFT_rel}
\hat{\alpha}_1=\frac{1}{2}\alpha_1\,,\qquad
\hat{\beta}_1=-\frac{b_1\Omega_{d-2}}{c_1 d\Omega_{d-1}}\,,\qquad
\hat{\alpha}_2=\left[\frac{1}{2}+\frac{b_1^2\Omega_{d-2}^2}{c_1c_2d(d-2)\Omega_{d-1}^2}\right]\alpha_2\,.
\end{equation}
The origin of the second term in $\hat{\alpha}_2$ is the correction term in \eqref{RmuConversion} in the relation between $\mu$ and $Q$.
As expected,  the scaling dimension of the lowest dimensional boundary operator is related to the one at the boundary as $\hat{\Delta}_Q\approx\frac{1}{2}\Delta_{2Q}$, where the equality holds up to $\mO(Q)$ corrections.  In particular, the difference between $\hat{\Delta}_Q-\frac12\Delta_{2Q}$ is controlled by the unique Wilson coefficients $b_1$ to leading order.  Notice also that the expansion in eq.~\eqref{eq_deltabdry} contains both even and odd powers of the cutoff $\mu\sim Q^{\frac{1}{d-1}}$, since parity is explicitly broken by the boundary.

We now quantize the system as before.  In particular, the phonon field $\pi(x)$ obeys the same eq.  \eqref{eq_EFT_lin_eq_bulk} in the bulk. However now the Neumann boundary condition $n^\mu \pd_\mu\pi\vert_{\pd\mathcal{M}}=0$ restricts the space of solutions only to the harmonics which are even under reflection across the equator. Their number in $d$ dimensions is given by~\cite{math}:
\begin{equation}\label{eq_multiplicity_even}
n^+_{\ell}=\frac{\Gamma (d-1+\ell)}{\Gamma (d-1) \Gamma (\ell+1)}\,.
\end{equation} 
The single particle states which form the Fock space have frequency $\omega_{\ell}$ as before.  However, since the boundary breaks the rotation group to $SO(d-1)$, the angular momentum of these states is found by decomposing the corresponding representation of $SO(d)$ into irrep.s of $SO(d-1)$. More precisely, the even states in the spin $\ell$ irrep of $SO(d)$ decompose precisely into the irreps $m=\ell,\ell-2,\ldots,-\ell$ of $SO(d-1)$. Therefore there is a huge accidental degeneracy between states with different quantum numbers. We expect that this degeneracy will be partially lifted upon including $1/\mu$ effects in the boundary conditions.

Finally the Casimir energy of the Fock states provides a one-loop contribution to the ground-state energy \eqref{eq_deltabdry}:
\begin{equation}
\delta\hat{\Delta}_Q^{(1-loop)}=\frac{1}{2}\sum_{\ell}n^+_{\ell}\omega_{\ell}\,.
\end{equation}
This correction provides a universal $\log Q$ correction to the boundary operator scaling dimension \eqref{eq_deltabdry}. Indeeed,
differently from eq.~\eqref{eq_bulk_Delta},  for large charge boundary operators in all (integer) dimensions there is either a boundary or a bulk operator which contributes at order $Q^0$ in the expansion in eq.~\eqref{eq_deltabdry}. Correspondingly the Casimir energy  should contain a logarithmic divergence generically. We indeed find such a divergence, and this leads to a universal $\log Q$ term in the large charge expansion of the scaling dimension of the boundary operator (in all integer dimensions $d$). To compute it, we regularize the calculation in a manner compatible with all the symmetries in dimensional regularization as in~\cite{Cuomo:2020rgt} and we find the result:\footnote{More physically, the logarithmic term can be extracted in any regularization scheme upon replacing the renormalization scale with the chemical potential $\mu$: for instance, in a cutoff scheme its coefficient agrees with $-1/(d-1)$ times the coefficient of the $\log\Lam$ term, where $\Lam$ is the energy cutoff, while in zeta-function regularization it is $-1/(d-1)$ times the coefficient of $\zeta(1)$. (The $-1/(d-1)$ factor comes from the conversion between $R\mu$ and $Q$ in \eqref{RmuConversion}.)}
\begin{equation}
\delta\hat{\Delta}_{Q}^{(1-loop)}=\hat{\gamma}_d \log Q+\mO\left(Q^{-\frac{1}{d-1}}\right)\,,
\end{equation}
where the coefficient $\hat{\gamma}_d$ depends on the number of spacetime dimensions:
\begin{equation}
\hat{\gamma}_{d}=\begin{cases}\displaystyle
-\frac{1}{64\sqrt{2}}\qquad  &\text{for }d=3\,,\\[8.5pt] \displaystyle
-\frac{1}{96\sqrt{3}}\qquad  &\text{for }d=4\,,\\[8.5pt] \displaystyle
-\frac{45}{8192 }\qquad  &\text{for }d=5\,,\\[8.5pt] \displaystyle
-\frac{1}{120\sqrt{5}}\qquad  &\text{for }d=6\,.
\end{cases}
\end{equation}
In $d=4$ and $d=6$ the coefficient $\hat{\gamma}_d$ is half of that which appears in the analogous contributions in the bulk scaling dimension in eq.  \eqref{EFTbulkQuantum}. This is because the operator renormalizing the Casimir energy in even dimensions is a bulk operator, whose coefficient is not affected by the presence of a boundary (the $1/2$ is because the integration is over half-space in the BCFT).

Finally we mention that, as explained in~\cite{MoninCFT}, the EFT also predicts correlation functions of light operators in between two large charge states.  For instance a scalar boundary operator of scaling dimension $\delta$ and charge $q$ can be matched in the EFT as:
\begin{equation}
\hat{\mO}_{q,\delta}=C(\hat{\pd}\chi)^\delta e^{iq \chi}+\ldots\,,
\end{equation}
where $C$ is an unknown Wilson coefficient. Proceeding as in~\cite{MoninCFT} we immediately find the following prediction for the OPE coefficient:
\begin{equation}
\hat\lambda_{q,\delta}=\langle Q+q|\hat{\mO}_{q,\delta}|Q\rangle\propto\left( \frac{2Q}{c_1}\right)^{\frac{\delta}{d-1}}\,.
\end{equation}
In particular we can also compute the OPE coefficient for the displacement operator in between two large charge operators. As explained in the introduction, this is obtained upon taking the boundary limit of the bulk stress tensor.  Working in hyperspherical coordinates,  this is given by:
\begin{equation}
\hat\lambda_{\hat D}=\langle Q|\hat{D}|Q\rangle=\lim_{\theta\rightarrow\frac{\pi}{2}}
\langle Q| T_{\theta\theta}(\theta)|Q\rangle=\frac{\alpha_1}{(d-1)\Omega_{d-1}}(2Q)^{\frac{d}{d-1}}\,,
\end{equation}
where $\alpha_1$ is defined in eq.~\eqref{eq_bulk_Delta}.

\section{A weakly coupled example: the \texorpdfstring{$O(2)$}{O(2)} BCFT in the \texorpdfstring{$4-\varepsilon$}{4-eps} expansion}
\label{sec:O2BCFTinEpsilonExpansion}

In this section we focus on the $O(2)$ model at the Wilson-Fisher fixed point in $d=4-\varepsilon$ dimensions on the  half-plane $x^d \geq 0$. 
Two boundary conditions are consistent with conformal invariance and also preserve the global $U(1)$ symmetry:\footnote{A third boundary condition, which defines the so-called extra-ordinary transition point~\cite{PhysRevB.12.3885}, is also compatible with conformal invariance, but breaks instead the $U(1)$ symmetry.}
\begin{enumerate}
\item Neumann boundary condition: $\pd_d\phi\vert_{x^d=0}=0$ .
\item Dirichlet boundary condition: $\phi\vert_{x^d=0}=0$ .
\end{enumerate}
We study the scaling dimensions $\hat{\Delta}_Q$ of the lowest dimension operator of charge $Q$ under the global $U(1)$ symmetry in the boundary theory for each of these boundary conditions.  This will allow us to verify explicitly many of the predictions discussed in the previous section.

\subsection{General considerations}

Consider the following (Euclidean) action in $d=4-\varepsilon$ dimensions in flat space:
\begin{equation}\label{eq_action}
S=\int d^dx\left(\left|\pd\phi\right|^2
+\frac{\lambda_0}{4}|\phi|^4\right),
\end{equation}
where $\phi$ is a complex scalar field, and $\lambda_0$ is the bare coupling constant, which to one-loop order is related to the physical one by the relation
\begin{equation}\label{eqCounterterm}
\lambda_0 M^{-\varepsilon}=\lambda+\frac{1}{\varepsilon}\delta\lambda_1+\frac{\delta\lambda_2}{\varepsilon^2}+\ldots \, ,\qquad
\delta\lambda_1=5\frac{\lambda^2}{(4\pi)^2}-\frac{15}{2} \frac{\lambda^3}{(4\pi)^4}+\mO\left(\frac{\lambda^4}{(4\pi)^6}\right)\, .
\end{equation}
Here $M$ is the sliding scale.
The $\beta$-function of $\lambda$ to two-loops order in perturbation theory is given by~\cite{Kleinert}
\begin{equation}\label{eq_beta_function}
\beta_\lambda=M\frac{d\lambda}{dM}=\lambda\left[
-\varepsilon+5\frac{\lambda}{(4\pi)^2}
-15\frac{\lambda^2}{(4\pi)^4}+\mO(\lambda^3)\right].
\end{equation}
This has a fixed-point at
\begin{equation}\label{lambdaStar}
\frac{\lambda_*}{(4\pi)^2}=\frac{\varepsilon }{5}+\frac{3 \varepsilon ^2}{25}+\mO(\varepsilon^3).
\end{equation}

We will focus the theory \eqref{eq_action} on the half-plane $x^d\geq 0$.  We are primarily interested in the scaling dimension of the lightest boundary operator of a given $U(1)$ charge $Q$. As emphasized in~\cite{Badel}, the diagrammatic perturbative expansion breaks down for correlators of charge $Q$ operators with $\lambda_* Q\sim \varepsilon Q\gtrsim 1$, due to the large combinatorial factors associated with multi-legged amplitudes. Instead we will work in the double scaling limit $\varepsilon\rightarrow 0$ with $\varepsilon Q = \text{fixed}$, for which the result takes the form:\footnote{A similar double scaling limit was analyzed in $\mathcal{N}=2$ superconfomal field theories~\cite{Bourget:2018obm,Beccaria:2018xxl,Grassi:2019txd} as well as in the large $N$ expansion~\cite{Alvarez-Gaume:2019biu,Giombi:2020enj}.}
\begin{equation}\label{eq_Delta_epsilon_gen}
\hat{\Delta}_Q=\frac{1}{\lambda_*}\hat{\Delta}_{-1}(\lambda_*Q)+\hat{\Delta}_0(\lambda_* Q)+\ldots
\end{equation} 
As we review below, the result in this limit is obtained expanding the path-integral around the appropriate semiclassical trajectory sourced by the operators insertions. Unsurprisingly, the saddle-point takes the form of a superfluid profile, and the parameter $\lambda_* Q$ controls the gap of the fluctuations of the radial mode.  In the regime of small $\lambda_* Q\ll (4\pi)^2$ the radial mode is light, and the coefficients in eq.~\eqref{eq_Delta_epsilon_gen} match the results of standard diagrammatic calculations in the vacuum.\footnote{For instance, in the $O(N)$ model with no boundary the result of the semiclassical calculation in the double scaling limit was succesfully compared with the outcome of a four loop diagrammatic calculation in~\cite{Jack:2021ypd}. } In the opposite regime $\lambda_* Q\gg (4\pi)^2$ the radial mode becomes heavy. Therefore it decouples, and we can compare $\hat{\Delta}_{Q}$ with the predictions discussed in sec.~\ref{Sec_EFT}.\footnote{Note that there are weakly coupled theories in which the gap
of the first excited state is controlled directly by $Q$ and thus the EFT applies for $Q\gg 1$ for any value of the coupling $\lambda$. For instance this is the case for monopole operators carrying $Q$ units of topological charge in $U(1)$ gauge theories with a large number $N_f$ of matter fields: in the monopole background, the spectrum of matter modes is organized in Landau levels, whose gap is controlled by the large magnetic field $B\sim Q$ directly, despite the existence of a small coupling constant $\lam=1/N_f$.  In these models the EFT is most naturally formulated in terms of the gauge field dual to the Goldstone field $\chi$. Computations of $\Delta_Q$ in these theories were performed in the UV description in~\cite{Murthy:1989ps,Borokhov:2002ib,Dyer:2013fja,Dyer:2015zha,delaFuente:2018qwv}.}

For future reference, we quote here the result for the scaling dimension $\Delta_Q$ of the bulk operator $\phi^Q$ obtained in~\cite{Badel} for $\varepsilon Q\gg 1$:
\begin{multline}\label{eq_Delta_epsilon_bulk}
\Delta_Q=\frac{1}{\varepsilon}\left(\frac{2}{5}\varepsilon Q\right)^{\frac{4-\varepsilon}{3-\varepsilon}}\left[\frac{15}{8}+\varepsilon\left(a_1+\frac38\right)+\mO\left(\varepsilon^2\right)\right]\\
+\frac{1}{\varepsilon}
\left(\frac{2}{5}\varepsilon Q\right)^{\frac{2-\varepsilon}{3-\varepsilon}}
\left[\frac{5}{4}-\varepsilon\left(a_2-\frac14\right)+\mO\left(\varepsilon^2\right)\right]+\ldots\,,
\end{multline}
where $a_1$ and $a_2$ are numerical constants:
\begin{equation}\label{eq_epsilon_bulk_coeff}
a_1=-0.5753315(3)\,,\qquad
a_2=-0.93715(9)\,.
\end{equation}
Comparing with eq.~\eqref{eq_bulk_Delta} and using eq.~\eqref{lambdaStar}, we then infer that the Wilson coefficients $c_1$ and $c_2$ of the bulk effective action \eqref{eq_action_LO_bulk} read:
\es{eq_epsilon_c}{
c_1&=
\frac{1}{\lambda_*}\left\{1+\frac{\lambda_*}{(4\pi)^2}\left[
\frac{5}{2} \left(\gamma_E -1+\log \pi \right)-8 a_1
\right]
+\mO\left(\frac{\lambda_*^2}{(4\pi)^4}\right)
\right\}\,,\\
c_2&=
\frac{1}{\lambda_*}\left\{\frac13+\frac{\lambda_*}{(4\pi)^2}\left[
\frac{5}{2}\left(\gamma_E+\log \pi\right)+\frac{17}{6}
-\frac{4}{3}\left(4 a_1+3 a_2\right)
\right]
+\mO\left(\frac{\lambda_*^2}{(4\pi)^4}\right)
\right\}
\,.
}
These expressions will be used to verify that the coefficients of the large $\lambda_* Q$ expansion of $\hat{\Delta}_Q$ satisfy the relations \eqref{eq_EFT_rel} predicted by the EFT.

To compute the coefficients in eq.~\eqref{eq_Delta_epsilon_gen}, we proceed as in~\cite{Badel}. Namely we consider the theory on $\mathds{R}\times\mathds{H}S^{d-1}$.  The action reads:
\begin{equation}
\int_{\theta\leq \pi/2} d^dx\sqrt{g}\left[|\pd\phi|^2+m_d^2|\phi|^2+\frac{\lambda_0}{4}|\phi|^4\right]\,,
\end{equation}
where $m_d$ is the conformal mass given by $m_d=\frac{d-2}{2 R}$ and  $R$ is the sphere radius.  Since it is the lowest energy eigenvalue in the charge $Q$ sector on the cylinder, the scaling dimension $\hat{\Delta}_Q$ can be extracted from the expectation value of the (Euclidean) evolution operator $e^{-HT}$ in between an arbitrary charge $Q$ state $|\psi_Q\rangle$ in the limit $T\rightarrow\infty$:
\begin{equation}\label{eq_epsilon_evolution}
\langle\psi_Q|e^{-HT}|\psi_Q\rangle\stackrel{T\rightarrow\infty}{=}\mathcal{N} e^{-T\hat{\Delta}_Q/R}\,.
\end{equation}
A convenient choice of the state is given by:
\es{trialPsi}{
|\psi_Q\rangle=\int\mathcal{D}\chi\exp\left[i\frac{Q}{2R^{d-1}\Omega_{d-1}}\int_{\theta\leq \frac{\pi}{2}} d\Omega\,\chi\right]|f,\chi\rangle\,,
}
where $|\rho,\chi\rangle$ denotes a state with fixed values of the field (in Schr\"odinger picture) in the polar parametrization $\phi=\frac{\rho}{\sqrt{2}}e^{i\chi}$. The choice of the value of $f=f(\theta)$ is in our hands and it will be used to simplify the calculations.  The path-integral corresponding to eq.~\eqref{eq_epsilon_evolution} may then be written as:
\begin{equation}
\langle\psi_Q|e^{-HT}|\psi_Q\rangle=\mathcal{Z}^{-1}\int_{\rho=f(\theta)}^{\rho=f(\theta)}\mathcal{D}\chi\mathcal{D}\rho \
e^{-S_{eff}}\,,
\end{equation}
where $S_{eff}$ is given by:
\begin{equation}\label{eq_epsilon_Seff}
S_{eff}=\int_{-T/2}^{T/2} d\tau\int_{\theta\leq\frac\pi2} d\Omega_{d-1}\left[
\frac12(\pd\rho)^2+\frac{1}{2}\rho^2(\pd\chi)^2+\frac{m_d^2}{2}\rho^2+\frac{\lambda_0}{16}\rho^4+i\frac{Q}{2R^{d-1}\Omega_{d-1}}\dot{\chi}
\right]\,.
\end{equation}
The factor $\mathcal{Z}$ ensures that the vacuum to vacuum amplitude is normalized to unity:
\begin{equation}
\mathcal{Z}= \int\mathcal{D}\chi\mathcal{D}\rho \
e^{-S}\,.
\end{equation}
Performing the path integral in the saddle point approximation then yields the result \eqref{eq_epsilon_evolution}. In particular the leading order arises from evaluating the action \eqref{eq_epsilon_Seff} on the solution of the equations of motion with boundary conditions specified by the trial wave function \eqref{trialPsi}. We choose $f(\theta)$ such that the saddle point configuration is stationary, i.e. 
\begin{equation}
\chi=-i\mu \tau\,,\qquad
\rho=f(\theta)\,,
\end{equation}
where $\mu$ and $f(\theta)$ are solutions of the following equations
\begin{equation}\label{eq_epsilon_EOMs}
\begin{aligned}
&\frac{\pd_\theta\left[\sin^{d-2}\theta\pd_\theta f(\theta)\right]}{R^2\sin^{d-2}\theta}+\left(\mu^2-m_d^2\right)f(\theta)-\frac{\lambda_0}{4}f^3(\theta)=0\, ,\\
&Q=R^{d-1}\int_{\theta\leq \frac{\pi}{2}} d\Omega_{d-1}\ \mu f^2(\theta)\,,
\end{aligned}
\end{equation}
supplemented by the condition $\pd_\theta f \vert_{\theta=\frac\pi2}=0$ for Neumann boundary conditions and by $f \vert_{\theta=\frac\pi2}=0$ for Dirichlet. Below we present the results for both cases.

\subsection{Neumann boundary conditions}

The Neumann boundary conditions are compatible with a constant profile for $\rho$. Therefore the equations of motion \eqref{eq_epsilon_EOMs} simplify:
\es{MuLamRel}{
\mu^2-m_d^2=\frac{\lambda_0}{4}f^2\,,\qquad
Q=\frac{1}{2}\Omega_{d-1} R^{d-1}\mu f^2\,,
}
where $\Omega_{d-1}$ is the volume of the unit  $(d-1)$-sphere. To leading order, eqs.~\eqref{MuLamRel} can be solved directly in four dimensions and yield the result:
\begin{equation}
\frac{1}{\lambda_*}\hat{\Delta}_{-1}(\lambda_* Q)=\frac{1}{2\lambda_*}\Delta^{bulk}_{-1}(2\lambda_* Q)\,,
\end{equation}
where $\Delta_{-1}^{bulk}(\lambda Q)/\lambda$ is the leading order result for the scaling dimension of the bulk operator $\phi^Q$ obtained in~\cite{Badel}. We do not report the exact expression here, but we just display the expansions for small and large $\lambda_* Q$:
\begin{equation}\label{eq_Delta1_neumann}
\frac{1}{\lambda_*}\hat{\Delta}_{-1}(\lambda_* Q)=\begin{cases}
\displaystyle
 Q \left[ 1 +    \frac{\lambda_* Q}{16\pi^2}
  + \mO\left(\frac{(\lambda_* Q)^2}{(4\pi)^4}\right) \right] , &\text{for}\quad \lambda_* Q \ll (4\pi)^2 \,,\\
\displaystyle \frac{4\pi^{2}}{\lambda_*} \left[ \frac{3}{4} \left( \frac{\lambda_* Q}{ 4\pi^{2}} \right) ^{4/3} + \frac{1}{2}\left( \frac{\lambda_* Q}{ 4\pi^{2}} \right) ^{2/3} + 
\mO \left(1\right) \right], &\text{for}\quad \lambda_* Q \gg (4\pi)^2.
\end{cases}
\end{equation}
The result for small $\lambda_* Q$ agrees with the diagrammatic result for the anomalous dimension $\hat{\gamma}_Q=\hat{\Delta}-Q\frac{d-2}{2}$ of the operator $\hat{\phi}^Q$, which to 1-loop is given by (see appendix \ref{appendix_diagrams} for details):
\begin{equation}\label{eq_Neumann_anomalous}
\hat{\gamma}_Q=\frac{\lambda_*}{16\pi^2}\left(Q^2-2Q\right)=\frac{\varepsilon}{5}\left(Q^2-2Q\right)\,.
\end{equation}
The large $\lambda_* Q$ result instead takes the pattern expected from the superfluid EFT. As in the example of sec.~\ref{SecInvitation}, integrating out $\rho$ at the classical level does not lead to any boundary term in the EFT in this case; correspondingly, this procedure does not produce any contribution linear in $Q$ in the large charge result and the relation \eqref{eq_DeltaQ_intro} is exact to all orders in the charge. This will not be true upon including quantum corrections, as it can already be seen from eq.~\eqref{eq_Neumann_anomalous}.

Let us now compute the one-loop correction to $\hat{\Delta}_Q$ in the double-scaling limit. This correction arises from the Casimir energy of the Goldstone and the radial modes:\footnote{The calculation in the rest of this subsection follows closely the general steps done in section 4 of~\cite{Badel}
.}
\begin{equation}
\frac{R}{2}\sum_{\ell=0}^\infty n^+_{\ell}\left[\omega_+(\ell)+\omega_-(\ell)\right]\,,
\end{equation}
where $n^+_{\ell}$ is the number of even hyperspherical modes on the sphere $S^{d-1}$ given in eq.~\eqref{eq_multiplicity_even} and the frequencies are given by:
\begin{equation}
\omega^2_\pm (\ell) = J^2_\ell+3\mu^2-m_d^2\pm \sqrt{4 J^2_\ell\mu^2+(3\mu^2-m_d^2)^2},\qquad J_{\ell}^2=\frac{\ell(\ell+d-2)}{R^2}\,.
\label{eq:dispersions}
\end{equation}
Using eq.~\eqref{eqCounterterm} to renormalize the divergence, we obtain the expression:
\begin{equation}\label{eqDelta0neumannPre}
\hat{\Delta}_0=\left\{\lim_{\varepsilon\rightarrow 0}\left[\frac{R}{2}\sum_{\ell=0}^{\infty}n^+_{\ell}\left[\omega_+(\ell)+\omega_-(\ell)\right]+\frac{5\left(\mu^2R^2-1\right)^2}{16\varepsilon}\right]\right\}_{\lambda_0=\lambda_*}\,.
\end{equation}
Performing the divergent part of the sum in dimensional regularization, the final result takes the form:
\begin{equation}\label{eqDelta0neumann}
\hat{\Delta}_0(\lambda_* Q)=-\frac{15 (\mu_*  R)^4+30 (\mu_*  R)^2-13}{32 }
+\frac{1}{2}\sum_{\ell=1}^{\infty}\sigma(\ell)+\frac{\sqrt{3\mu_*^2 R^2-1}}{\sqrt{2}}
\end{equation}
where we defined
\begin{equation}\label{sigell}
\sigma(\ell)=\frac{1}{2} (\ell+1) (\ell+2)R\left[\omega^*_+(\ell)+\omega^*_-(\ell)\right]
-\ell^3-4 \ell^2-\ell \left(\mu_*^2 R^2+4\right)-2 \mu_*^2 R^2+\frac{5 \left(\mu_*^2 R^2-1\right)^2}{8 \ell}\,.
\end{equation}
Here $*$ stresses that all quantities are evaluated in $d=4$ with $\lambda_0$ equal to the fixed point value $\lambda_*$ (which enters expressions through \eqref{MuLamRel}). The sum over $\sigma(\ell)$ cannot be evaluated in closed form in general, but it is convergent since $\sigma(\ell)\sim 1/\ell^3$.

Let us now consider the result in the regime of small $\lambda_* Q$. 
In this case we can expand $\sigma(\ell)$ in powers of $\lambda_* Q$ and then perform the sum in eq.~\eqref{eqDelta0neumann} analytically order by order. Adding then the tree-level result \eqref{eq_Delta1_neumann} to the so evaluated one-loop contribution \eqref{eqDelta0neumann}, and using \eqref{lambdaStar}, we find the following anomalous dimension:
\begin{equation}
\hat{\gamma}_Q=\varepsilon\left(\frac{Q^2}{5}-\frac{2 Q}{5}\right)  -
\varepsilon^2\left[\frac{2 Q^3}{25}-\left(\frac{8}{25}-\frac{2 \pi ^2}{75}\right) Q^2\right] +\mO\left(\varepsilon^3Q,\varepsilon^3 Q^4\right)\,.
\end{equation}
This clearly agrees with the diagrammatic result in equation \eqref{eq_Neumann_anomalous}. 

We can also evaluate the Casimir energy in the regime $\lambda_* Q\gg (4\pi)^2$. One option is to follow the methodology of~\cite{Badel}, which led to the result quoted in \eqref{eq_epsilon_bulk_coeff}. That procedure involves obtaining terms that contain a $ \log \left( \lambda_* Q \right)$ enhancement analytically, and then numerically evaluating the remaining sum over $\sig(\ell)$ from \eqref{sigell}, and determining coefficients in the large  $\lambda_* Q$ expansion from fitting to these numerical values. In appendix \ref{App_Casimir_epsilon}, we present an improvement over this method that is analytic and hence does not involve fitting. It leads to the final result:
\begin{multline}\label{eq_epsilon_Delta0_large}
\hat{\Delta}_0=\left[\frac{5 }{48} \log \left( \frac{\lambda_* Q}{ 4\pi^{2}} \right) 
+\hat{a}_1\right]\left( \frac{\lambda_* Q}{ 4\pi^{2}} \right)^{4/3} 
+\hat{d}_1 \left( \frac{\lambda_* Q}{ 4\pi^{2}} \right) \\
+\left[-\frac{5 }{72} \log \left( \frac{\lambda_* Q}{ 4\pi^{2}} \right) 
+\hat{a}_2\right]\left( \frac{\lambda_* Q}{ 4\pi^{2}} \right)^{2/3}
+\hat{d}_2 \left( \frac{\lambda_* Q}{ 4\pi^{2}} \right)^{1/3}+\mO\left(1\right)\,.
\end{multline}
The coefficients $\hat{a}_i,\, \hat{d}_i$ are determined by definite integrals as detailed in appendix \ref{App_Casimir_epsilon}, e.g.
\es{a1expr}{
\hat{a}_1= {\frac{5}{32}}\le[2\ga_E-3-\log\le(\frac{8}{5}\ri)\ri]+\frac12\int_{0}^\infty dk \ \widehat{\Sigma}_1(k)\,,
}
where $\widehat{\Sigma}_1(k)$ is given in \eqref{SigExp} and \eqref{SigHat}. The coefficients can be easily evaluated numerically to arbitrary precision:
\begin{align}\label{eq_epsilon_Neumann_coeff}
\hat{a}_1=-0.287665\,,\qquad
\hat{d}_1=-0.530918\,,\qquad
\hat{a}_2=-0.468560\,,\qquad
\hat{d}_2=0.173701\,.
\end{align}
The result is in agreement with the general structure \eqref{eq_deltabdry}. To see this, it is convenient to add eq.~\eqref{eq_epsilon_Delta0_large} to the leading order \eqref{eq_Delta1_neumann} and write the result in the form:
\begin{equation}
\begin{split}
\hat{\Delta}_Q&=\frac{1}{\varepsilon}\left(\frac{4}{5}\varepsilon Q\right)^{\frac{d}{d-1}}\left[\frac{15}{16}+\varepsilon\left(\hat{a}_1+\frac{3}{16}\right)+\mO\left(\varepsilon^2\right)\right]+\frac{1}{\varepsilon}
\left(\frac{4}{5}\varepsilon Q\right)\left[0+\varepsilon\hat{d}_1
+\mO\left(\varepsilon^2\right)\right]\\
&+\frac{1}{\varepsilon}\left(\frac{4}{5}\varepsilon Q\right)^{\frac{d-2}{d-1}}\left[\frac{5}{8}+\varepsilon\left(\hat{a}_2-\frac{1}{8}\right)+\mO\left(\varepsilon^2\right)\right]
+\frac{1}{\varepsilon}\left(\frac{4}{5}\varepsilon Q\right)^{\frac{d-3}{d-1}}
\left[0+\varepsilon\hat{d}_2+\mO\left(\varepsilon^2\right)\right]
+\ldots
\end{split}
\end{equation}
We may now also use this expression to verify the relations \eqref{eq_EFT_rel}. In particular, since $b_1=0$ at tree-level (corresponding to $\mO(1/\varepsilon)$),  the EFT demands that the coefficients of the $Q^{\frac{d}{d-1}}$ and $Q^{\frac{d-2}{d-1}}$ are precisely half of those in eq.~\eqref{eq_Delta_epsilon_bulk} for the bulk scaling dimension. This implies $\hat{a}_1=a_1/2$ and $\hat{a}_2=a_2/2$, in beautiful agreement with the values in eqs.~\eqref{eq_epsilon_bulk_coeff} and \eqref{eq_epsilon_Neumann_coeff}.\footnote{With the improved technology of appendix \ref{App_Casimir_epsilon}, we can give an analytic proof of these relations and also determine the coefficients $a_1$ and $a_2$ with arbitrary precision.} Furthermore, using the leading order value of $c_1$ given in eq.~\eqref{eq_epsilon_c}, we find that the coefficient $b_1$ of the boundary action \eqref{eq_action_bdry_LO} is given by:
\begin{equation}
b_1=-\frac{\hat{d}_1}{2\pi}+\mO\left(\frac{\lambda_*}{(4\pi)^2}\right)\,.
\end{equation}
Therefore a positive  coefficient $b_1$ is generated at the quantum level (at $O(\ep^0)$), corresponding to an increase in the charge density close to the boundary.

\subsection{Dirichlet boundary conditions}

The Dirichlet boundary conditions are clearly not compatible with a non-trivial constant profile. Instead below we solve the eqs.~\eqref{eq_epsilon_EOMs} for small and large values of $\lambda Q$. We will only work at leading order in the double scaling limit, therefore we can set $d=4$ in the following.

For small $\lambda Q$,  the lowest dimension operator of charge $Q$ corresponds to $(\pd_d\hat{\phi})^Q$, whose anomalous dimension $\hat{\gamma}_Q = \hat{\Delta}_Q -Q\frac{d}{2}$ is computed in appendix \ref{appendix_diagrams} and reads:
\begin{equation}\label{eq_anomalous_Dir}
\hat{\gamma}_Q=\frac{\lambda_*}{32\pi^2}\left(Q^2-3 Q\right)+\mO\left(\frac{\lambda_*^2Q^3}{(4\pi)^4}\right)\,.
\end{equation}
We can reproduce the leading $\mO\left(\lambda_* Q^2\right)$ term of this result with a semiclassical calculation. To this aim we notice that for small $\lambda_* Q/(4\pi)^2$ the non-linear term in eq.~\eqref{eq_epsilon_EOMs} can be neglected to leading order. Treating it perturbatively, we found the solution up to order $\mO\left((\lambda_*Q)^6\right)$. The leading orders read:
\es{fProf}{
f(\theta) &=\frac{\sqrt{2Q}}{\pi R}\cos\theta\left[1+\frac{2\lambda_* Q}{384\pi^2}-\frac{2\lambda_* Q}{48\pi^2}\cos^2\theta+\mO\left(\frac{\lambda_*^2 Q^2}{(4\pi)^4}\right)\right]\,,\\[7pt]
R\mu &=2+\frac{2\lambda_* Q}{32\pi^2}+\mO\left(\frac{\lambda_*^2 Q^2}{(4\pi)^4}\right)\,.
}
Using the classical profile (to higher order than displayed above) to evaluate the action \eqref{eq_epsilon_Seff}, we find the result:
\begin{equation}\label{eq:DirichletSmallCouplingDelta}
\begin{split}
\frac{1}{\lambda_*}\hat{\Delta}_{-1}(\lambda_* Q)=Q&\left[ 2+\frac{\lambda_* Q}{2(4\pi)^2}-\frac{\lambda_*^2 Q^2}{3(4\pi)^4}+
\frac{31}{72}\frac{ \lambda_*^3 Q^3}{(4 \pi) ^6}
-\frac{2491 }{3456}\frac{\lambda_*^4 Q^4}{(4\pi)^8}\right.\\
&\left.
+\frac{57763 }{41472 }\frac{\lambda_*^5 Q^5}{(4\pi)^{10}}
-\frac{14686201 }{4976640}
\frac{\lambda_*^6 Q^6}{ (4\pi) ^{12}}
+\mO\left(\le(\frac{\lambda_* Q}{(4\pi)^2}\ri)^7\right)\right]\,,
\end{split}
\end{equation}
whose first two terms clearly agree with eq.~\eqref{eq_anomalous_Dir}.

In the large $\lambda_* Q$ regime we can use the method of matched asymptotic expansions in solving the saddle point equations~\eqref{eq_epsilon_EOMs}. Let us define the variable $x\equiv\cos\theta$, and as in
sec.~\ref{SecInvitation} define the far region with $x=\mO(1)$ and the near boundary region where the rescaled coordinate $X\equiv \mu R x=\mO(1)$ (and hence $x\to 0$ as $\mu R \to \infty$). The solution in the far region  can be found in an $\exp(- \sqrt{2} \mu R x)$ expansion (but exactly in $\mu$):
\es{farSol}{
f(x)={\frac{2\sqrt{(\mu R)^2-1}}{ \sqrt{\lam}}}\le[1+{\cal O}\le(\exp(-\sqrt{2} \mu R x)\ri)\ri]
}
In the near region we solve the equation in the $1/(\mu R)$ expansion. From the condition of no singularity at finite $X$ as well as the Dirichlet boundary condition, we get a unique profile at leading order in $\mu R$:
\es{nearSol}{
f(X)=\frac{2\mu}{\sqrt{\lambda}}\tanh\left(\frac{X}{\sqrt{2}}\right)\left[1+\mO\left(\frac{1}{(\mu R)^2}\right)\right] \,.
}
Free coefficients arise at subleading orders.
The two series can be matched in their overlapping regime of validity, for $X\to\infty$ (but $x\to 0$). The terms we have written explicitly match at leading order, and the procedure fixes the undetermined coefficients at subleading orders. We have performed the matching explicitly to ${\cal O}\le(e^{-\sqrt{2}X}\ri)$ and to ${\cal O}\le(1/\mu\ri)$, but will not write the explicit formulas here.

We may now use this solution to compute the charge and the classical action as a function of $\mu$. Combining those expressions we find the final result for the scaling dimension:
\begin{equation}\label{eq_epsilon_Dirichlet_large}
\begin{split}
\frac{\hat{\Delta}_{-1}(\lambda_* Q)}{\lambda_*}=&
\frac{4\pi^2}{\lambda_*}\left[
\frac{3}{4} \left( \frac{\lambda_* Q}{ 4\pi^{2}} \right)^{4/3}+
\frac{4\sqrt{2}}{3 \pi}\left(\frac{\lambda_* Q}{4\pi^2}\right)
\right.\\[8pt]&\left.
+
\frac{\left(32+3 \pi ^2\right)}{6\pi^2}
\left(\frac{\lambda_* Q}{4\pi^2}\right)^{2/3}+
\mO\left(\left(\frac{\lambda_* Q}{4\pi^2}\right)^{1/3}\right)
\right]\,.
\end{split}
\end{equation}
As expected, comparing with eq.~\eqref{eq_Delta1_neumann},  the coefficient of the leading term $\sim(\lambda_* Q)^{4/3}$ is the same for both Neumann and Dirichlet boundary conditions. For Dirichlet we find a positive classical
term linear in $Q$, corresponding to a negative coefficient $b_1$ for the EFT boundary action \eqref{eq_action_bdry_LO}:
\begin{equation}\label{eq_Neumann_b1}
b_1=
-\frac{4 \sqrt{2}}{3 \lambda_* }\left[1+\mO\left(\frac{\lambda_*}{(4\pi)^2}\right)\right]
\,.
\end{equation}
As in the example of sec.~\ref{SecInvitation}, this coefficient is needed to match in the EFT the decrease in the charge density close to the boundary. Finally,  we notice that the $Q^{2/3}$ term in the EFT is determined from $b_1$, $c_1$ and $c_2$ by the relations \eqref{eq_EFT_rel}.  We may then use the explicit values of the $c_i$'s in eq.~\eqref{eq_epsilon_c} for this model, together with the value of $b_1$ in eq.~\eqref{eq_Neumann_b1}, to compare the EFT prediction with the result for the $Q^{2/3}$ term in eq.~\eqref{eq_epsilon_Dirichlet_large}. We find perfect agreement, providing an additional non-trivial check of the EFT approach.

\section{Other large charge phases in BCFTs}\label{SecOther}

Large charge operators in CFTs are not always described by a superfluid EFT. Alternative phases are for instance found in free theories, $\mathcal{N}\geq 2$ SCFTs~\cite{Hellerman:2017veg,Hellerman:2017sur,Hellerman:2018xpi} and free fermions~\cite{Komargodski:2021zzy}. Similarly, there also exist BCFTs with different large charge phases. Here we discuss some examples.

\subsection{Free charged scalar with interacting boundary}

The large charge sector of CFTs with moduli, such as free theories or $\mathcal{N}\geq 2 $ SCFTs in four dimensions, behaves differently than in generic theories. This is because the corresponding EFT is formulated in terms of an axio-dilaton complex scalar field $\phi$ with flat potential. As a result the lowest dimensional charged operator obeys $\Delta_Q\propto Q$ for $Q\rightarrow\infty$~\cite{Hellerman:2017veg}.  

Here we analyze what happens when coupling such theories to a boundary, which may partially lift the flat direction, focusing on the example of a free bulk theory. A possibility which is special to BCFT is to couple free bulk theories to interacting boundary degrees of freedom, see e.g.~\cite{Herzog:2017xha,DiPietro:2019hqe} for perturbative examples.  Here we consider the case where the free bulk scalar is charged under a $U(1)$ symmetry preserved by the boundary.  

We consider first a simple toy model, given by:
\begin{equation}\label{eq_free_model2}
\int d^dx|\pd\phi|^2+\frac{\lambda}{4}\int d^{d-1}x|\phi|^4\,,
\end{equation}
where the boundary conditions are perturbation of Neumann ones.
This has a fixed point in $3-\varepsilon$ dimensions at the zero of the beta function
$\beta_{\lambda}=-\varepsilon\lambda+5\frac{\lambda^2}{4\pi}+\mO\left(\frac{\lambda^2}{(4\pi)^2}\right)$. We would like to study the scaling dimension of the lowest dimensional operators with large $U(1)$ charge $Q\sim 1/\lambda_*$. To this aim, we proceed as in sec.~\ref{Sec_EFT}. Namely we consider the theory on the strip $\mathds{R}\times\mathds{H}S^{d-1}$ and expand the field around a profile of the form:
\begin{equation}
\phi=e^{-i\mu t} f(\theta)\,,
\end{equation}
where we work in spherical coordinates and the function $f$ solves the equations of motion.  To leading order we work in $d=3$, for which the bulk equation reads
\begin{equation}\label{eq_sec5_EOMs}
\frac{1}{R^2\sin\theta}\pd_\theta\left[\sin\theta \pd_\theta f\left(\theta\right)\right]+
\left(\mu^2-\frac{1}{4R^2}\right)f\left(\theta\right)=0\,,
\end{equation}
while the boundary condition and the condition of fixed charge imply
\begin{equation}\label{eq_sec5_bdry_cond}
\left[\pd_\theta f(\theta)/R+ \frac{\lambda}{2}f^3(\theta)\right]_{\theta=\frac{\pi}{2}}=0\,,\qquad
Q=2\mu R^2 \int_{\theta\leq\frac{\pi}{2}}d\Omega_2 f^2(\theta)\,.
\end{equation}

To solve the eqs.~\eqref{eq_sec5_EOMs} and \eqref{eq_sec5_bdry_cond}, we notice that, for arbitrary $\mu$, the regular solution of the bulk equation can be written in terms of a hypergeometric function:
\begin{gather}
f(\theta)=v \,\, _2F_1\left(\frac{1+2 R\mu }{2},\frac{1-2 R\mu }{2};1;\frac{1-\cos\theta}{2}\right)\,.
\end{gather}
To find the value of $v$ and $\mu$ we may instead plug this expression in the eq.s~\eqref{eq_sec5_bdry_cond}, and solve them as a function of the charge and the coupling. Though we were not able to find a solution in closed form for general values of $Q$, it is possible to check numerically that such a solution always exists,  and that the chemical potential  and $v$ satisfy $\frac{1}{2R}\leq \mu \leq \frac{3}{2R}$ and $v^2\propto Q$.  In the following we discuss the explicit result for small and large $\lambda Q$.

At small $\lambda Q$, the solution is close to the one describing unperturbed Neumann, $\phi=v e^{-i\frac{t}{2R}}$.  Therefore,  expanding the chemical potential around $\mu\simeq\frac{1}{2R}$ we find 
\begin{equation}
\mu=\frac{1}{2R}+\frac{\lambda Q}{4\pi R}+\mO\left(\frac{\lambda^2Q^2}{(4\pi)^2R}\right)\,,\qquad
v=\sqrt{\frac{Q}{2\pi R}}\left[1-
\frac{\lambda Q}{8\pi}\left(1+\log 2\right)
+\mO\left(\frac{\lambda^2Q^2}{(4\pi)^2R}\right)
\right]\,.
\end{equation}
Computing the classical energy we find
\begin{equation}\label{eq_sec5_DeltaQ_moduli}
\hat{\Delta}_Q=\frac{Q}{2}+\frac{\lambda Q^2}{8\pi}+\mO\left(\frac{\lambda^2Q^3}{(4\pi)^2}\right)\,,
\end{equation}
which is in perfect agreement with the diagrammatic result for the anomalous dimension of the boundary operator $\hat{\phi}^Q$,
\begin{equation}
\hat{\gamma}_Q=\frac{\lambda Q(Q-1)}{8\pi}+\mO\left(\frac{\lambda^2 Q^3}{(4\pi)^2}\right)\stackrel{\lambda=\lambda_*}{=}\frac{\varepsilon Q(Q-1)}{10}+
\mO\left(\varepsilon^2 Q^3\right)
\,.
\end{equation}
 
Let us now consider the large $\lambda Q$ regime. In this case the absence of a bulk potential implies that the chemical potential stays of order one, differently than in the $O(2)$ model studied in sec.~\ref{sec:O2BCFTinEpsilonExpansion}.  Nonetheless the boundary interaction implies that the coefficient of the ratio $\hat{\Delta}_Q/Q$ changes compared to eq.~\eqref{eq_sec5_DeltaQ_moduli}. To see this we notice that $\lambda v^2\sim \lambda Q$. Therefore in the limit $\lambda Q\rightarrow\infty$, the boundary condition \eqref{eq_sec5_bdry_cond} demands that $f(\theta)/v$ approaches zero. This is \emph{effectively} analogous to a Dirichlet condition.  In this limit $\mu\simeq \frac{3}{2}$ so that $f(\theta)\simeq v\cos\theta$ and we find that the scaling dimension reads:
\begin{equation}
\hat{\Delta}_Q=\frac{3}{2}Q+\mO\left(\frac{(\lambda Q)^{2/3}}{\lambda}\right)\,.
\end{equation}

Consider now a more general theory in $d$ dimensions, in which additional boundary degrees of freedom are coupled to the free field. In a large charge state we expect that these will be gapped by the large expectation value of the scalar field.\footnote{E.g. in the previous model one has $\lambda|\phi|^4\vert_{bdry}\sim  (\lambda Q)^{2/3}/\lambda$ for large $\lambda Q$.} Integrating them out, we will then produce a potential $\sim|\phi|^{\frac{2(d-1)}{d-2}}$ in the boundary, and we expect a description similar to the model we just discussed to apply. In particular the energy of the ground state should coincide with the free Dirichlet answer $\hat{\Delta}_Q\simeq\frac{d}{2}Q$ to leading order in $Q$.  

We also expect that similar considerations apply to the lowest dimensional large charge operators for more general bulk theories that have moduli, e.g. in four-dimensional $\mathcal{N}\geq 2$ SCFTs coupled to a superconformal boundary,  especially when the boundary conditions break enough supersymmetry. \footnote{Superconformal boundary conditions in $\mathcal{N}\geq 2$ superconformal theories in four dimensions were analyzed, e.g.,  in~\cite{Erdmenger:2002ex,Gaiotto:2008sa,Gaiotto:2008ak}.}

\subsection{Free fermion in four dimensions}

As a next example, let us consider a free massless Dirac fermion in four dimensions. In component it reads:
\begin{equation}
\psi_D \equiv \begin{pmatrix} \psi_\alpha \\ \zeta^\dagger_{\dot{\alpha}} \end{pmatrix}= \begin{pmatrix} \psi_1 \\ \psi_2 \\ \zeta^\dagger_1 \\ \zeta^\dagger_2 \end{pmatrix} \,.
\end{equation}
In the absence of a boundary,  the free Dirac field enjoys a $U(1)\times U(1)$ symmetry under which the phases of $\psi_\alpha$ and $\zeta^\dagger_{\dot{\alpha}}$ shift independently. Bulk operators with charge $Q\gg 1$ under the diagonal $U(1)$ acting as $\psi_D\rightarrow e^{i\alpha}\psi_D$ were constructed explicitly in~\cite{Komargodski:2021zzy}. These correspond to Fermi spheres with all spinor harmonic levels filled up to spin $j=j_{max}-1$ and a number $\delta Q$ of modes filled in the $j=j_{max}$ level.  When the numbers of fermions $\delta Q$ in the last level vanishes, the scaling dimension reads:
\begin{equation}
\Delta_Q=\frac{3}{4}\left(\frac{3}{2}\right)^{\frac{1}{3}}Q^{\frac{4}{3}} + \frac{1}{2\cdot 2^{\frac{2}{3}}\cdot 3^{\frac{1}{3}}}
Q^{\frac{2}{3}}+\mO\left(Q^{-\frac{2}{3}}\right).
\end{equation}
When some fermions are present in the last level, the $Q^{4/3}$ terms is unchanged, but the subleading corrections are different.

Now we consider the theory in the half plane $x_3 \geq 0$.  The are two conformal boundary conditions. In components, these differ only by a sign and read:
\begin{equation}
\left.\left( \psi_1\pm\zeta^\dagger_1 \right)\right\vert_{\text{bry}} = 0\quad\text{and}\quad
 \left.\left( \psi_2\mp\zeta^\dagger_2 \right)\right\vert_{\text{bry}} = 0\,.
 \end{equation}
Therefore, it is clear that adding a boundary to the free Dirac fermionic theory reduces the number of total independent fermionic degrees of freedom by half.  
The internal symmetry is further broken to a single $U(1)$. We conclude that, for the boundary CFT,  the scaling dimension of the lightest operator of charge $Q$ under the unbroken $U(1)$ is exactly related to the bulk scaling dimension of the operator with charge $2Q$: 
\begin{equation}
\hat{\Delta}_Q = \frac{1}{2}\Delta_{2Q}\,.
\end{equation}

\subsection{Theories with charged degrees of freedom only at the boundary}

It is possible to consider models in which all charged states are made of boundary degrees of freedom, i.e. in which the Noether current is a boundary operator and there is no bulk current. For these models, large charge states are clearly not described by the EFT discussed in this paper. Here we provide a few general comments.

Let us first gain some intuition through discussing an example on the half-plane. Namely we consider the following classical model in four spacetime dimensions
\begin{equation}
S=\int_{x\geq 0} d^4x\left[\frac{1}{2}(\pd\phi)^2-\frac{\lambda}{4}\phi^4\right]
+\int_{z=0} d^3x\left[|\pd\psi|^2-g_1|\psi|^2\phi^2+\frac{g_2}{2}|\psi|^4\phi-\frac{g_3}{3}|\psi|^6\right]\,,
\end{equation}
where $\phi$ and $\psi$ are, respectively, a real and a complex scalar field.  The boundary conditions for $\phi$ are perturbation of Neumann and break explicitly the $\mathds{Z}_2$ symmetry of the bulk action.  We assume all couplings to be positive.  For sufficiently large $g_3$, the classical solution with finite charge density is given by:
\begin{equation}
\psi= e^{i\mu t}v\,,\qquad
\phi=\sqrt{\frac{2}{\lambda}}\,\frac{1}{z+c/\mu}\,,
\end{equation}
where $v^2\sim \mu/\lambda$ and $c\sim\mO(1)$, the precise value being determined by the equation of motion for $\psi$ and the boundary condition for $\phi$. We see that the chemical potential therefore not only gaps the boundary radial mode of $\psi$, but it also sources a non-trivial one-point function for the bulk field. In particular in the $\mu\rightarrow\infty $ limit the boundary conditions for the bulk field are exactly the same as those defining the \emph{extra-ordinary} transition at leading order in the epsilon expansion~\cite{Dey:2020lwp}. Therefore the low energy EFT describing this model is given by a boundary superfluid coupled to the (irrelevant) boundary operators of the $\phi^4$ theory with the extra-ordinary boundary conditions. In this case the only primary boundary operator is the displacement operator $\hat{D}$~\cite{Liendo:2012hy} (see appendix~\ref{subsec:BackgroundFieldsApproach} for a review), which has dimension $4$, and the EFT reads: 
\begin{equation}
S_{EFT}=S_{\text{extra-ordinary}}+c\int_{z=0} d^3x (\hat{\pd}\chi)^3+
c_{\hat{D}}\int_{z=0} d^3x\hat{D}(\hat{\pd}\chi)^{-1}+\ldots\,.
\end{equation}

For more general theories we expect something similar to happen. Namely the presence of a chemical potential in the boundary might drastically change the boundary conditions for the bulk fields, and therefore the boundary spectrum in the $\mu\rightarrow\infty$ limit. Notice that in general the boundary conditions will break boosts and therefore will not define a BCFT sector even in the $\mu\rightarrow \infty $ limit.  However, coupling this  scale invariant sector (representing the bulk degrees of freedom) to the boundary superfluid through irrelevant boundary operators nonlinearly realizes the symmetries of the microscopic BCFT, and hence should provide a complete description of its physics at energies $E\ll \mu$.

Large charge operators on the strip should admit a similar description. In particular the charge density can only accumulate at the boundary and the energy of large charge operators with minimal scaling dimension scales as:
\begin{equation}
\hat{\Delta}_Q\propto Q^{\frac{d-1}{d-2}}\,.
\end{equation}
We further generically expect a subsector of the theory to be described by a superfluid Goldstone boson on the boundary. However in this case also some \emph{bulk} degrees of freedom will remain gapless, and to determine the full spectrum of charged BCFT operators we have to solve this generically strongly coupled sector of our effective theory.

\section{Other applications}\label{SecOther2}

\subsection{The large charge sector of defect CFTs}

It is simple to extend our ideas to general defect CFTs, i.e. CFTs in the presence of a conformal defect. Consider for instance a $p$-dimensional conformal defect or interface in a $d$-dimensional CFT with $U(1)$ symmetry. We expect large charge defect operators to be described by an EFT analogous to the one described in sec.~\ref{Sec_EFT}, with the same bulk action and a defect action generically scaling as $\sim\mu^p\sim Q^{\frac{p}{d-1}}$.\footnote{For $p=d-2$,  the EFT is analogous to that describing vortices in superfluids~\cite{Horn:2015zna}, the only difference being that the latter are dynamical and therefore describe new states in the \emph{bulk} CFT - see~\cite{Cuomo:2017vzg}.}
This implies that the scaling dimension of large charge defect operators differs from that of bulk operators only at order $Q^{\frac{p}{d-1}}$:\footnote{More generally, this scaling only depends on dimensional analysis and not on the specific form of the EFT.}
\begin{equation}
\Delta_Q^{(defect)}-\Delta_Q^{(bulk)}\sim Q^{\frac{p}{d-1}}.
\end{equation}
Notice that in the case of an interface between two identical theories $p=d-1$, and the difference scales linearly with the charge analogously to the corrections to eq.~\eqref{eq_DeltaQ_intro} for BCFTs. Also the spectrum of excited states is largely unchanged, being described by a Fock space of phonons with the same dispersion relation to leading order in $\mu$.

\subsection{Thermodynamics and OPE coefficients in BCFTs}

It is expected that, in non-integrable CFTs, OPE coefficients of a light operator $\mO$ in between two heavy operators $H$ and $H'$ are controlled by the Eigenstate Thermalization Hypothesis (ETH)~\cite{Lashkari:2016vgj}.  In this section we review the relevant statements and provide their generalization to BCFTs.

 Consider first diagonal matrix elements of the form $\langle H|\mO|H\rangle$. In the limit $\Delta_H\rightarrow\infty$ the equivalence between microcanonical and canonical ensembles implies that these coincide with the thermal expectation value $\langle\mO\rangle_{\beta}\simeq b_{\mathcal{O}} 
\beta^{-\Delta_{\mO}}$, where $b_{\mathcal{O}}$ is a numerical coefficient which depends on the operator. As it follows from simple dimensional analysis, the temperature in this equivalence is parametrically set by the energy density $\epsilon_H\sim \Delta_H/R^{d}$ of the state $|H\rangle$ as $\epsilon_H\simeq\frac{d-1}{d}b_T/\beta^d$~\cite{Delacretaz:2020nit}, where $b_T$ is the coefficient in front of the thermal one-point function of the stress tensor and it is thus proportional to the $C_T$ central charge of the theory (see e.g.~\cite{Iliesiu:2018fao} for details on thermal correlators in CFT). Therefore we conclude:
\begin{equation}\label{eq_preETH}
\langle H|\mO|H\rangle\simeq \langle\mO\rangle_{\beta}\sim \Delta_H^{\frac{\Delta_{\mO}}{d}}\,.
\end{equation}
Eq.~\eqref{eq_preETH} equivalently follows from the requirement that the correlator obeys the macroscopic limit~\cite{Lashkari:2016vgj,Jafferis:2017zna}.

The extension of eq.~\eqref{eq_preETH} to off-diagonal matrix elements is provided by the Eigenstate Thermalization Hypothesis (ETH) ansatz, that states
\begin{equation}\label{eq_ETH}
\langle H|\mO|H'\rangle=\delta_{H\,H'}\langle\mO\rangle_{\beta}+
\Omega^{-1/2}\left(\frac{\Delta_H+\Delta_{H'}}{2}\right) R^{\mathcal{O}}_{H H'}\,,
\end{equation}
where $\Omega\left(\Delta\right)$ is the density of states with energy $\Delta$ and $R^{\mathcal{O}}_{H H'}$ are random variables, whose variance is set by the four point-function $\langle H|\mO\mO|H\rangle$ and does not scale exponentially with $\Delta_H$ and $\Delta_{H'}$~\cite{Delacretaz:2018cfk,Delacretaz:2020nit}. In the thermodynamic limit the density of states scales as:
\begin{equation}\label{eq_ETH_entropy}
\Omega(\Delta_H)\sim e^{S (\Delta_H)}\,,\qquad
S(\Delta_H)\propto \Delta_H^{\frac{d-1}{d}}\,,
\end{equation}
where $S(\Delta_H)$ is the entropy of the system
at a temperature $\beta^{-1}\sim \Delta_H^{{1/d}}$ and the powers are again dictated by dimensional analysis (for the entropy density $S(\Delta_H)/R^{d-1}$).

Let us now consider BCFTs. From the viewpoint of the state-operator correspondence it is natural to expect a relation of the form \eqref{eq_ETH} for matrix elements of \emph{both} boundary and bulk operators in between two heavy boundary states. Due to the locality of the theory,  the relation between temperature and energy density is unmodified to leading order. We may further use locality to argue that the value of the entropy of the system with a boundary $\hat{S}(\Delta_H)$, and therefore of the density of states, admits a relation similar to eq.~\eqref{eq_DeltaQ_intro} with the entropy of the bulk CFT:
\begin{equation}\label{eq_ETH_entropy2}
\hat{S}(\hat\Delta_H)\simeq \frac{1}{2}S(2\hat\Delta_H)\,.
\end{equation}
The factor of two in the argument of $S$ follows from the fact that the temperature in the thermodynamic limit is set by the energy density and not the total energy. The $1/2$ in front instead follows because, by locality, the leading contribution to the partition function on the hemisphere depends only on the total volume. As for eq.~\eqref{eq_DeltaQ_intro}, we expect that corrections will scale as the area of the boundary, and therefore in the thermodynamic limit will be suppressed by a relative factor $\Delta_H^{-{1/d}}$.

\subsection{Solid EFT near a boundary}\label{SecSolid}

It is clear that the EFT approach towards boundary conditions is independent of conformal symmetry and might therefore be used to study more general 
setups.  In particular, the EFT approach allows to account for the non-linear realization of the spontaneously broken spacetime symmetries in the spirit of \cite{Nicolis:2015sra}. As an illustration, here we construct the most general boundary conditions for the phonon modes in a solid close to a boundary.

Let us first recall the construction of the EFT for a homogeneous and isotropic solid in infinite volume \cite{Dubovsky:2005xd}. 
From the low energy viewpoint, a solid can be defined as a theory invariant under the Poincar\'e group and an internal group $G$ isomorphic to the $d-1$-dimensional Euclidean group, in which boosts and the spatial Euclidean group $E_{d-1}$ are broken to the diagonal group of $G$ and $E_{d-1}$ \cite{Nicolis:2013lma}.  Intuitively, $G$ is an emergent symmetry which
accounts for the regularity of the crystal structure, whose symmetry group is approximately continuous in the long wavelength limit. The most economic way to realize this symmetry breaking pattern is to introduce $d-1$ scalars $\phi^I$ that under the internal group $G$ transform as:
\begin{equation}
\phi^I\rightarrow R^I_{\;J}\phi^J+a^J\,,\qquad
R\in O(d-1)\,.
\end{equation}
The fields are expanded around the expectation value
\begin{equation}\label{eq_SolidBB}
\langle \phi^I\rangle=\alpha x^I+\text{const.}\,,
\end{equation}
where $\alpha$ physically represents the compressibility of the solid. To leading order in derivatives the action is most conveniently written in terms of the following matrix
\begin{equation}
B^{IJ}=\pd_\mu \phi^I\pd^\mu \phi^J\,.
\end{equation}
Assuming parity, a complete set of $G$-invariants is given by:
\begin{equation}
A=\text{Tr}\left[B\right]\,,\quad
C_2=\frac{\text{Tr}\left[B^2\right]}{\text{Tr}\left[B\right]^2}\,,\quad\ldots\,,\quad
C_{d-1}=\frac{\text{Tr}\left[B^{d-1}\right]}{\text{Tr}\left[B\right]^{d-1}}\,.
\end{equation}
The bulk action is finally written as
\begin{equation}\label{eq_SolidAction}
S=\int d^dx F\left[A,C_2,\ldots ,C_{d-2}\right]\,,
\end{equation}
where $F$ is an arbitrary function.  

The physical meaning of eq.~\eqref{eq_SolidAction} becomes clear upon expanding into fluctuations $\pi^i=\alpha^{-1}\phi^i-x^i$, where the $\pi^i$ denote the $d-1$ phonon modes of the solid. One finds the following quadratic action:
\begin{equation}\label{eq_SolidActionQuadraticBulk}
S^{(2)}=-\frac{AF_A}{d-1}\int d^dx\left[\dot{\pi}^i\dot{\pi}^i-c_T^2(\pd_i\pi^j)^2-\left(c_L^2-c_T^2\right)(\pd_i\pi^i)^2\right]\,,
\end{equation}
where, $c_L$ and $c_T$ are the sound speed of the longitudinal and transverse modes respectively, whose value follow depends on the function $F$ in eq.~\eqref{eq_SolidAction}.\footnote{Explicitly these are given by \cite{Esposito:2017qpj}:
\begin{align}
c_T^2&=1+\frac{(d-1)}{F_A A}\sum_{n=2}^{d-1}\frac{\pd F}{\pd C_n}\frac{n(n-1)}{(d-1)^n}
\,,\\
c_T^2&=
1+\frac{2 F_{AA}A^2}{(d-1)F_A A}+\frac{2(d-2)}{F_A A}\sum_{n=2}^{d-1}\frac{\pd F}{\pd C_n}\frac{n(n-1)}{(d-1)^n}
\,,
\end{align}
where we denoted with subscripts derivatives with respect to $A$ and all quantities are evaluated on the background configurations $\pd_i\phi^J=\alpha \delta_i^J$. If we further assume conformal invariance, we must have $F=A^{d}f\left(C_2,\ldots,C_{d-1}\right)$ and the sound speeds are related as $c_L^2=\frac{1}{d-1}+2\frac{d-2}{d-1}c_T^2$.} Eq.~\eqref{eq_SolidActionQuadraticBoundary} therefore just states that the phonons have a linear dispersion relation in the long wavelength limit, as well known.  The advantage of this construction is that Lorentz invariance constrains the interaction of these modes upon further expanding eq.~\eqref{eq_SolidAction}.

Let us now consider the solid near its a boundary in the $n$ direction at $x^n=0$.  Focusing momentarily only on the linearly realized symmetries, a boundary breaks translations in the $n$th direction. Therefore the shift symmetry of the $\pi^{N}$ phonon associated with the breaking of translations in the $n$th direction must be explcitly broken by the boundary conditions. Rotations and translations in the parallel direction should instead be preserved by a boundary at $x^n=0$. To leading order in derivatives, a natural guess for such boundary conditions reads:
\begin{equation}\label{eq_solid_BC_guess}
\pd_n\pi^i\vert_{x^n=0}=0\quad i\neq N\,,\qquad
\pi^N\vert_{x^n=0}=0\,,
\end{equation}
corresponding to Neumann boundary conditions for the phonons propagating parallel to the boundary, and Dirichlet for the orthogonal one. In the following we shall see how to obtain eq.~\eqref{eq_solid_BC_guess} as well as its first derivative corrections within the EFT framework outlined above.

Notice first that the boundary corresponds to the endpoint of the lattice structure. Therefore we expect that the boundary not only breaks the spacetime symmetry, but also part of the internal group $G$. Let us call $\phi^A$ with $A=1,\ldots,d-2$ the $d-2$ fields whose background expectation value is proportional to the coordinates parallel to the boundary, and $\phi^N$ the remaining one. We expect that the boundary preserves only the subgroup $H\subset G$ corresponding to shifts and internal rotations of the fields $\phi^A$.  We therefore consider the most general boundary conditions compatible with this group.

As in the superfluid case, the shift invariance of the fields $\phi^A$ implies that their boundary conditions may be thought as perturbations of Neumann ones.\footnote{This is because charge conservation implies that the bulk currents corresponding to the shift symmetry obey $J_A^n=-\hat{\pd}_a \hat{J}^a_A$ at the boundary,  where $\hat{J}_A$ is a boundary current, which is one-derivative suppressed - see e.g. Appendix \ref{subsec:BoundaryVSBulkCurrentAndEMT}.  Consistently with this observation, the unbroken diagonal rotations imply $J_I^n\propto \delta_I^n$ on the background.} The boundary condition for $\phi^N$ will instead generically be of the mixed type, and to leading order in derivatives it will be written in terms of a non-linear function of $\pd_n\phi^N$ and $\phi^N$ itself at $x^n=0$.  On the background $\pd_\mu\phi^N=\alpha\delta_\mu^n$ such a boundary condition specifies the value of $\phi^N$ at the boundary, breaking explicitly the shift symmetry. As for the superfluid, we may discuss systematically this state of affairs introducing the most general boundary action for the fields $\phi^I$ and their derivatives parallel to the boundary, such that the group $H$ is preserved.

To build the boundary action, we define a matrix $\hat{B}^{AB}$ as follows:
\begin{equation}
\hat{B}^{AB}=\hat{\pd}_\mu \phi^A\hat{\pd}^\mu \phi^B\,,
\end{equation}
where $\hat{\pd}_\mu=\{\pd_0,\pd_a\}$ denotes the derivative along the coordinates parallel to the boundary.
Out of traces of $\hat{B}$ we can build the following $H$-invariants:
\begin{equation}
\hat{A}=\text{Tr}\left[\hat{B}\right]\,,\quad
\hat{C}_2=\frac{\text{Tr}\left[\hat{B}^2\right]}{\text{Tr}\left[B\right]^2}\,,\quad\ldots\,,\quad
\hat{C}_{d-2}=\frac{\text{Tr}\left[\hat{B}^{d-2}\right]}{\text{Tr}\left[B\right]^{d-1}}\,.
\end{equation}
Finally we notice that the field $\phi^N$ is $H$-invariant, and therefore the boundary action may depend on it as well. This also implies that we can systematically neglect derivatives of $\phi^N$ in the boundary action to the first non-trivial order. Overall, the most general boundary action reads:\footnote{For a conformal solid we would have $G=\hat{A}^{d-1}g(\phi^N,\hat{C}_2,\ldots)$. }
\begin{equation}\label{eq_SolidBoundaryAction}
S_{bdry}=\int_{x^n=0} d^{d-1}x\,G\left[\phi^N,\hat{A},\hat{C}_2,\ldots
,\hat{C}_{d-2}\right]\,.
\end{equation}

Varying eq.~\eqref{eq_SolidAction} and eq.~\eqref{eq_SolidBoundaryAction} we may now see that the background profile \eqref{eq_SolidBB} is largely unaffected:
\begin{equation}\label{eq_Solid_profile}
\phi^A=\alpha x^a\delta_a^A+\text{const}.\,,\qquad
\phi^N=\alpha x^n\delta_n^N+c^N\,.
\end{equation}
The constant term for the $\phi^A$ fields is unspecified as before. The only difference with eq.~\eqref{eq_SolidBB} is that the constant contribution $c^N$ cannot be shifted arbitrarily and instead follows from the boundary condition
\begin{equation}
\left[\frac{\pd F}{\pd\pd_n\phi^N}-\frac{\pd G}{\pd \phi^N}\right]_{x^n=0}=
-\left[\frac{2}{d}F_A A+\frac{\pd G}{\pd \phi^N}\right]_{x^n=0}
=0\,,
\end{equation}
where all quantities are evaluated on the profile \eqref{eq_Solid_profile}. For instance a boundary action $G=-m^{d-1}\left(\phi^N\right)^2+\ldots$ with $m\rightarrow\infty$ dictates $\phi^N=0$ corresponding to Dirichlet boundary conditions.\footnote{It is possible to write the boundary action in an equivalent \emph{dual} form, which is manifestly smooth for pure Dirichlet conditions. We explain how to do this for a simple example in the appendix \ref{App_Dual_Bactions}.} In general the equilibrium value $c^N$ depends also on the compressibility $\alpha$ of the solid through the functions $F$ and $G$. In practice we do not need to solve explicitly for the $c^N$.

As before, the physical meaning of this construction is appreciated once we expand in fluctuations. The boundary action for the phonon modes read:\footnote{In writing the bulk action \eqref{eq_SolidActionQuadraticBulk} we integrated by parts discarding some boundary terms. These however only renormalize those in eq.~\eqref{eq_SolidActionQuadraticBoundary} and therefore do not affect our analysis.}
\begin{equation}\label{eq_SolidActionQuadraticBoundary}
\begin{split}
S_{bdry}^{(2)}=-\frac{G_A \hat{A}}{(d-2)}\int_{x^d=0} d^{d-1}x &
\left[\dot{\pi}^a\dot{\pi}^a-
\hat{c}_T^2(\pd_a\pi^b)^2-\left(\hat{c}_L^2-\hat{c}_T^2\right)(\pd_a\pi^a)^2
\right.\\ &\left.
+2\mu_{N}\pi^N\pd_a\pi^a-m_N^2(\pi^N)^2
\right]\,,
\end{split}
\end{equation}
where the explicit value of the coefficients is obtained expanding the function $G$ in eq.~\eqref{eq_SolidBoundaryAction}.\footnote{The explicit expressions are given by:
\begin{align}
&\hat{c}_T^2=1+\frac{(d-2)}{G_A \hat{A}}\sum_{n=2}^{d-2}\frac{\pd G}{\pd \hat{C}_n}\frac{n(n-1)}{(d-2)^n}
\,,\\
&\hat{c}_T^2=
1+\frac{2 G_{AA}\hat{A}^2}{(d-2)G_A \hat{A}}+\frac{2(d-3)}{G_A \hat{A}}\sum_{n=2}^{d-2}\frac{\pd G}{\pd \hat{C}_n}\frac{n(n-1)}{(d-2)^n}
\,,\\
&\mu_N=-\alpha\frac{\pd G_A/\pd\phi^N}{G_A}\,,\qquad
m_N^2=\frac{\pd^2G/\pd(\phi^N)^2}{2 G_A}\,.
\end{align}}
Eq.~\eqref{eq_SolidActionQuadraticBoundary} is manifestly invariant under the unbroken rotations, and in fact it could have been guessed without keeping track of the full broken group.  However, the nonlinear terms that we neglected are instead constrained by Lorentz invariance, and may be straightforwardly analyzed upon expanding the boundary action to higher orders.  Notice that the terms in eq.~\eqref{eq_SolidActionQuadraticBoundary} are all of the same order in the derivative expansion since, as we will see explicitly below, $\pi^N\sim \pd\pi$ because of the boundary conditions.

From eqs. \eqref{eq_SolidActionQuadraticBulk} and \eqref{eq_SolidActionQuadraticBoundary} we finally obtain the boundary conditions for the phonon modes. These may be written as:
\begin{align}\nonumber
&c_T^2\pd_n\pi^a\vert_{x^n=0}=\frac{G_A\hat{A}(d-1)}{F_A A(d-2)}\left[\ddot{\pi}^a-\hat{c}_T^2\pd_b\pd^b\pi^a-\left(\hat{c}_L^2-\hat{c}_T^2\right)\pd_a\pd_b\pi^b+\mu_N\pd_a\pi^N\right]_{x^n=0}\,,\\[2ex]
&
m^2_N\pi^N\vert_{x^n=0}=\left\{\mu_N\pd_a\pi^a+
\frac{F_A A(d-2)}{G_A\hat{A}(d-1)}\left[c_T^2\pd_n\pi^N+\left(c_L^2-c_T^2\right)\pd_i\pi^i\right]\right\}_{x^n=0}\,.
\label{eq_SolidBcondition}
\end{align}
Neglecting the right-hand side, eqs.~\eqref{eq_SolidBcondition} reduce to the guess \eqref{eq_solid_BC_guess}. Corrections are one derivative suppressed and are clearly compatible with the unbroken symmetry group. Our construction additionally shows that these are also compatible with the underlying Poincar{\'e} symmetry.

\section*{Acknowledgements}

We thank M.~Metlitski and L.~Rastelli for useful discussions.  
We are grateful to Z.~Komargodski for 
collaboration at the early stages of this project and
valuable comments on a preliminary draft of this paper.
GC is supported by the Simons Foundation (Simons Collaboration on the Non-perturbative Bootstrap) grants 488647 and 397411. MM and ARM are supported in part by the Simons Foundation grant 488657 (Simons Collaboration on the Non-Perturbative Bootstrap) and the BSF grant no. 2018204. The work of ARM was also supported in part by the Zuckerman-CHE STEM Leadership Program.

\appendix

\section{Ward identities in BCFTs }\label{AppendixWardIdentity}

In this appendix we clarify some important properties of the energy-momentum tensor and the conserved currents in boundary (and defect) conformal field theories. In appendix \ref{subsec:BackgroundFieldsApproach}, we study the Ward identities associated with diffeomorphisms, Weyl invariance and internal symmetries by taking variations of the curved-spacetime effective action with respect to the background fields. In appendix \ref{subsec:BoundaryVSBulkCurrentAndEMT}, we study the consequences of the Nother procedure in the case of boundary conformal field theories. We comment on a certain puzzle that arises in perturbative calculations in these models, clarify its source and resolution.

\subsection{Background field approach to Ward identities in BCFTs}\label{subsec:BackgroundFieldsApproach}

In this appendix we derive the Ward identities associated with diffeomorphisms and Weyl invariance in BCFTs. We will show that diffeomorphism and Weyl invariance are always saturated by the bulk stress tensor and the displacement operator only, and we will derive the corresponding Ward identities. This is not entirely trivial because, \emph{a priori}, the variations of the metric and its normal derivatives may source additional operators on the boundary; however we will argue that these are excluded by the unitarity bounds in BCFTs. Our derivation provides an extension of the one given in~\cite{Jensen:2015swa}, where these additional contributions were not considered. 

We mostly use the notation and conventions of~\cite{Jensen:2015swa}. In particular, we  use the embedding formalism to parametrize the submanifold geometry~\cite{Aharony:2013ipa}. Let us consider a $d$ dimensional Riemannian manifold $\mathcal{M}$ equipped with a metric $g_{\mu\nu}$. We define the following:
\begin{itemize}
\item $x^\mu$, $\mu= 1,\cdots d$ denotes the bulk coordinates. 
\item $\sigma^a$, $a=1, \cdots d-1$ denotes the boundary coordinates. 
\item The embedding functions $X^\mu(\sigma^a)$ represent the boundary's position. 
\item The induced metric associated with the boundary is given by $\hat{g}_{ab} \equiv g_{\mu\nu}\partial_a X^\mu \partial_b X^\nu$. 
\item The bulk covariant derivative $\nabla^\mu$, with standard torsionless Levi-Civita connection associated with the background metric $g_{\mu\nu}$. The covariant derivative of a vector $A^\nu$ is thus given by $\nabla_\mu A^\nu \equiv \partial_\mu A^\nu + \Gamma^\nu_{\mu\rho}A^\rho$, and the Christoffel symbols are related to the metric through $\Gamma^{\nu}_{\mu\rho} = \frac{1}{2}g^{\nu\alpha}\left[-\partial_\alpha g_{\mu\rho}+\partial_\mu g_{\alpha\rho}+\partial_\rho g_{\alpha\nu} \right]$.
\item $\hat\nabla^a$ is the induced boundary covariant derivative, it acts on a mixed-index tensor $A^\mu_b$ in the following way:  
\begin{equation}
\hat\nabla_a A^\mu_b \equiv \partial_a A^\mu_b + \Gamma^\mu_{\nu a}A^\nu_b -\hat{\Gamma}^c_{ab}A^\mu_c,
\end{equation}
where $\hat\Gamma^c_{ab}$ is the Levi-Civita connection associated with the induced metric $\hat{g}_{ab}$, and $\Gamma^\mu_{\nu a}$ is the pullback of the Levi-Civita connection, defined by $\Gamma^\mu_{\nu a} \equiv \Gamma^\mu_{\nu\rho}\partial_a X^\rho$.
\item The projector tangential to the boundary submanifold is $P^\mu_\nu = g_{\nu\rho} \hat{g}^{ab}\partial_a X^\mu \partial_b X^\rho$. 
\item $n_\mu$ is a unit normalized ($n_\mu n^\mu =1$) foliation 1-form normal to the boundary submanifold. The following relations hold: $ g^{\mu\nu} = n^\mu n^\nu +P^{\mu\nu}$,  $\delta^\mu_\nu = n^\mu n_\nu +P^\mu_\nu$. Using the above definitions it is clear that $n^\nu P^\mu_\nu =n^\mu P_{\mu\nu} =0$.
\item We also define the second fundamental form: $\Pi^\mu_{ab} \equiv \hat{\nabla}_a \partial_b X^\mu$. Note that it is symmetric $\Pi^\mu_{ab} = \Pi^\mu_{ba}$ and satisfies $P^\mu_\nu \Pi^\nu_{ab} = 0$. The extrinsic curvature reads $K_{ab} = -n_\mu \Pi^\mu_{ab}$ (in the embedding formalism this is automatically symmetric $K_{ab}= K_{ba}$). We denote its corresponding trace by $K \equiv \hat{g}^{ab}K_{ab}$.
\end{itemize}

Let $Z=Z(g_{\mu\nu}, X^\mu)$ be the partition function of the theory as a function of the geometry.
The effective action $W = -\log Z(g_{\mu\nu}, X^\mu)$ is the generating functional of all the connected correlation functions in the theory.  Its variation with respect to the metric $g_{\mu\nu}$ and the embedding $X^\mu$ \emph{formally} define the energy momentum tensor and the displacement operator:
\begin{equation}
\label{eq:DefStressTensorOnePoint}
\langle T^{\mu\nu}_{\text{tot}}(x) \rangle \equiv -\frac{2}{\sqrt{g}} \left.\frac{\delta W}{\delta g_{\mu\nu}(x)}\right\vert_{X^\mu = \text{fixed}} \, ,
\end{equation}
\begin{equation}
\label{eq:DefDispOnePoint}
\langle \hat D_{\mu}(\sigma^a) \rangle \equiv -\frac{1}{\sqrt{\hat{g}}} \left.\frac{\delta W}{\delta X^{\mu}(\sigma^a)}\right\vert_{g_{\mu\nu} = \text{fixed}} \, ,
\end{equation}
where $g$ and $\hat{g}$ are the determinants of $g_{\mu\nu}$ and $\hat{g}_{ab}$ respectively and the displacement operator has support only at the boundary. At a closer look however eq.~\eqref{eq:DefStressTensorOnePoint} in general does not define a unique scaling operator, but rather a combination of bulk and boundary ones. Indeed in general the variation of the effective action with respect to $g_{\mu\nu}$ and $X^\mu$ receives contributions from both the bulk and the boundary. The most general form of $\delta W$ reads:
\begin{equation}\label{eq:VariationW}
\begin{aligned}
&\delta W   = -\frac{1}{2}\int_{\mathcal{M}} d^d x \sqrt{g} \, \delta g_{\mu\nu} \langle T^{\mu\nu}\rangle+
\int_{\partial{\mathcal{M}}} d^{d-1}\sigma \sqrt{\hat{g}}\delta X^\mu\langle \hat D_\mu\rangle \\
& \qquad   -\int_{\partial{\mathcal{M}}} d^{d-1}\sigma \sqrt{\hat{g}} \, \left[\frac{1}{2}\delta g_{\mu\nu} \langle \hat{T}^{\mu\nu}\rangle  + n^\rho\nabla_\rho \delta g_{\mu\nu}\langle \hat{A}^{\mu\nu} \rangle + n^\rho n^\sigma \nabla_\rho\nabla_\sigma \delta g_{\mu\nu} \langle \hat{B}^{\mu\nu}  \rangle +\cdots \right] 
\end{aligned}
\end{equation}
where the dots stand for higher-order normal derivatives acting on the metric.  The definition eq.~\eqref{eq:DefStressTensorOnePoint} then implies that $T^{\mu\nu}_{\text{tot}}$ receives contributions from several operators at the boundary:\footnote{The delta function in the direction normal to the boundary is defined in a general coordinate independent way as $\delta(x_\bot)\equiv\int_{\partial \mathcal{M}} d^{d-1}\sigma\, \sqrt{\hat g} \, \delta^d\left(x-X(\sigma)\right)/\sqrt{g}$. }
\begin{equation}\label{eqAppTmunu} 
T^{\mu\nu}_{\text{tot}} = T^{\mu\nu}+ \delta(x_{\bot})\hat{T}^{\mu\nu} - 2K \hat{A}^{\mu\nu}\delta(x_\bot)-2n^\alpha\nabla_\alpha\delta(x_\bot)\hat{A}^{\mu\nu} +\cdots    
\end{equation} 
where $K$ is the trace of the extrinsic curvature defined above.  Here $T^{\mu\nu}$ is the bulk stress tensor and $\hat{T}^{\mu\nu}$ the boundary stress tensor. The additional terms in the second line of eq.~\eqref{eq:VariationW} have no clear physical interpretations and are often neglected in the BCFT literature. We will argue below that indeed they must vanish in BCFTs.

We remark here that an alternative approach was previously discussed in the literature~\cite{McAvity:1993ue} (see also~\cite{Billo:2016cpy}), in which the boundary contribution to the variation \eqref{eq:VariationW} is not written in terms of normal derivatives of the metric, but in terms of the variation of an arbitrary number of geometric invariants, such as the induced metric, the extrinsic curvature, etc. While that approach is ultimately equivalent to ours,  we believe that the parametrization in eq.~\eqref{eq:VariationW} is more convenient, since it makes manifest which are the independent operators that may generically be sourced by geometric perturbations in BCFTs.  

First we notice that the sum is not infinite and it is restricted by dimensional analysis, and the unitarity bounds. In the following we retain all terms up to $\hat{A}^{\mu\nu}$ for simplicity of the presentation, but our results do not depend on this restriction.

Next we write down the Ward identities associated with reparametrization, diffeomorphism and Weyl invariance. These state that the effective action $W$ must be invariant under the following reparametrizations:
\begin{enumerate}
\item Reparametrization of the boundary coordinates, generated by a vector field $\zeta^a$: 
\begin{equation}
\delta_\zeta x^\mu =0, \qquad \delta_\zeta\sigma^a =-\zeta^a.
\end{equation} 
The metric and embedding functions transform according to:
\begin{equation}\label{eq_repa_var}
\delta_\zeta g_{\mu\nu} = 0 , \qquad \delta_\zeta X^\mu = \zeta^a\partial_a X^\mu. 
\end{equation}
\item Reparametrization of the bulk coordinates (diffeomorphisms), generated by a vector field $\xi^\mu$: 
\begin{equation}
\delta_\xi x^\mu = -\xi^\mu, \qquad \delta_\xi \sigma^a = 0.
\end{equation}
The metric and embedding functions change according to:
\begin{equation}\label{eq_diffeo_var}
\delta_\xi g_{\mu\nu} = \nabla_\mu\xi_\nu+\nabla_\nu\xi_\mu, \qquad \delta_\xi X^\mu = -\xi^\mu. 
\end{equation}
\item Weyl rescaling of the metric:
\begin{equation}
\delta_\Omega g_{\mu\nu} = 2\Omega(x)g_{\mu\nu}, \qquad \delta_\Omega X^\mu =0.
\end{equation}
\end{enumerate}
Obviously, the first two requirements hold for general boundary QFTs, while the last is special of BCFTs. In the following we will study the consequences of these requirements on the partition function of the theory.

Reparametrization of the boundary coordinates $\delta_\zeta W = 0$ implies:
\begin{equation}
\delta_\zeta W = \int d^{d-1} \sigma \sqrt{\hat{g}} \, \zeta^a\partial_a X^\mu \langle \hat D_\mu\rangle \,.
\end{equation}
This should vanish for an arbitrary $\zeta^a$, hence:
\begin{equation}
\label{eq:WardIdentityReparaBdry}
\partial_a X^\mu  \hat D_\mu =0 \,. 
\end{equation}
The above Ward identity states that all the components of $\hat D_\mu $ parallel to the boundary must vanish. The identity eq.~\eqref{eq:WardIdentityReparaBdry}, as well as those we will derive below, hold in correlation functions at separeted points.

Reparametrization of the bulk coordinates leads to the following requirement: 
\begin{equation}
\begin{aligned}
\label{eq:VariationExplicitlyDiff2}
0 &= -\frac{1}{2}\int_{\mathcal{M}} d^dx \sqrt{g} \, \delta_{\xi} g_{\mu\nu} \langle T^{\mu\nu} \rangle\\
&\quad + \int_{\partial{\mathcal{M}}} d^{d-1}\sigma \sqrt{\hat{g}}\, \left[ 
\delta_{\xi}X^\mu \langle \hat D_\mu \rangle-
\frac{1}{2}\delta_{\xi}g_{\mu\nu}\langle \hat{T}^{\mu\nu}\rangle - n^\rho\nabla_\rho \delta_\xi g_{\mu\nu}\langle \hat{A}^{\mu\nu}\rangle+ \cdots\right]\\
& = \int_{\mathcal{M}} d^dx \sqrt{g}\,\xi_\nu\nabla_\mu \langle T^{\mu\nu} \rangle  \\
&+ \int_{\partial\mathcal{M}} d^{d-1}\sigma \sqrt{\hat{g}}\, \left[n_\mu\xi_\nu \langle T^{\mu\nu}\rangle\vert_{\partial\mathcal{M}}
-\xi_\nu \langle \hat D^\nu \rangle
-\nabla_\mu\xi_\nu \langle \hat{T}^{\mu\nu} \rangle - n^\rho\nabla_{\rho}\nabla_{\mu}\xi_\nu \langle \hat{A}^{\mu\nu}\rangle +\cdots \right] ,
\end{aligned}
\end{equation}
where $\langle \hat D^\mu \rangle \equiv g^{\mu\nu} \langle \hat D_\nu \rangle$ and we integrated by parts in the second line; this picks a contribution from the bulk stress tensor at the boundary.
Therefore, the bulk stress-tensor remains conserved even in the presence of a boundary:
\begin{equation}
\nabla_\mu \langle T^{\mu\nu} \rangle = 0\,.
\end{equation}
The remaining terms should vanish for an arbitrary $\xi_\nu$.  To study the implications of this fact, let us consider first a small diffeomeorphism around the flat metric, in coordinates such that the boudary is at $x^d=0$.
In this case, equation \eqref{eq:VariationExplicitlyDiff2}  translates into the following expression:
\begin{equation}\label{eq:variationWardFlat}
\begin{aligned}
0 & = \int_{x^d=0} d^{d-1}x \, \left[ \xi_a \left( \langle T^{da}\rangle\vert_{x^d=0} +\partial_b \langle \hat{T}^{ab}\rangle - \langle \hat D^a \rangle \right) \right. \\
&\qquad \qquad \quad \quad \, \left. +\xi_d \left(\langle T^{dd}\rangle\vert_{x^d=0}+\partial_a \langle \hat{T}^{da}\rangle - \langle \hat D^d \rangle   \right) \right. \\
&\qquad \qquad \quad \quad \, \left. + \partial_d \xi_a \left( -\langle \hat{T}^{da}\rangle +\partial_b \langle \hat{A}^{ab} \rangle \right)\right. \\
&\qquad \qquad \quad \quad \, \left. + \partial_d \xi_d\left(-\langle \hat{T}^{dd}\rangle +\partial_a\langle \hat{A}^{da}\rangle \right)  \right]\\
&\qquad \qquad \quad \quad \, \left.  -\partial_d^2 \xi_a \langle \hat{A}^{da}\rangle -\partial_d^2 \xi_d \langle \hat{A}^{dd}\rangle \right] +\cdots \,.
\end{aligned}
\end{equation}
Since all normal derivatives of the vector $\xi_\mu$ are independent,  each parenthesis in eq.~\eqref{eq:variationWardFlat} should vanish.  Generalizing eq.~\eqref{eq:variationWardFlat} to an arbitrary curved manifold and using eq.~\eqref{eq:WardIdentityReparaBdry}, we find:\footnote{Here $\mathcal{R}_{\mu\nu\rho\sigma}$ is the Riemann tensor defined as $[\nabla_\rho,\nabla_\mu]v_\nu=-\mathcal{R}^{\sigma}_{\;\nu\rho\mu}v_\sigma$.}
\begin{align} \nonumber
& T^{na}\vert_{\pd\mathcal{M}}+\hat{\nabla}_bT^{ba}
+K^a_{\;b}\hat{T}^{bn}
-\hat{\nabla}_c\left(K^c_{\;b}\hat{A}^{ab}\right)-
\mathcal{R}_{nbc}^{\quad a}\hat{A}^{bc}=0\,,
\\
\nonumber
& T^{nn}\vert_{\pd\mathcal{M}}-\hat D^n-K_{ab}\hat{T}^{ab}+\hat{\nabla}_a\hat{T}^{an}+\left(K_a^{\;c}K_{cb}+\mathcal{R}_{nanb}\right)\hat{A}^{ab}=0\,,\\ 
& \hat{\nabla}_a\hat{A}^{ab}-\hat{T}^{nb}=0\,,
\label{eq_diffeoID}
\\ \nonumber
& \hat{T}^{nn}+K_{ab}\hat{A}^{ab}=0\,,\\
& \hat{A}^{nn}=\hat{A}^{na}=0\nonumber\,,
\end{align}
where we decomposed all tensors into transverse and parallel component using:
\begin{equation}
v^\mu=\pd_a X^\mu v^a+n^\mu v^n\,,\qquad v^n=v_n\,.
\end{equation}
The first eq. in \eqref{eq_diffeoID} in flat space reduces to $\partial_b \langle \hat{T}^{ab}\rangle=-\langle T^{da}\rangle\vert_{x^d=0} $ in agreement with Noether's theorem for translations along the boundary. However the identities \eqref{eq_diffeoID}, without further input, are not enough to rule out the existence of the operator $ \hat{A}^{ab}$ and of the component $\hat{T}^{da}$ of the boundary stress tensor, which have no clear physical interpretation.

Finally, ignoring trace anomalies, Weyl invariance requires:
\begin{equation}\label{eqWeylApp}
g_{\mu\nu} \langle T^{\mu\nu}_{\text{tot}} \rangle = 0, \qquad \text{(up to Weyl anomalies)}.
\end{equation}
Using eq.~\eqref{eqAppTmunu} this implies that $T^{\mu\nu}$, $\hat{T}^{\mu\nu}$, $\hat{A}^{\mu\nu}$, etc. are traceless.

We now explain how to rule out the existence of the additional operators in eq.~\eqref{eq:VariationW} in BCFTs in flat space, at least in low enough dimensions. Consider first $\hat{A}^{\mu\nu}$. Eqs.~\eqref{eq_diffeoID} set $\hat{A}^{dd}=\hat{A}^{ad}=0$.  The remaining component $\hat{A}^{ab}$ is a traceless symmetric operator of dimension $d-2$. By the unitarity bounds, which demand that any primary operator with spin $\ell$ satisfies $\hat\Delta_{\ell}\geq d-1-\ell$, we deduce that it cannot be a primary. The only other possibility is that it is a level $2$ descendant of a scalar, but this again is not compatible with the unitarity bound $\hat\Delta_{0}\geq\frac{d-3}{2}$ for $d< 5$.\footnote{In $d=5$ $\hat{A}^{ab}$ could be a level $2$ descendant of a free scalar and would therefore drop from eq.~\eqref{eq:variationWardFlat} by the free equations of motion. In $d>5$ a non-zero $\hat{A}^{ab}$, and therefore a non-vanishing $\hat{T}^{ad}=\pd_b\hat{A}^{ab}$, would be compatible with the vanishing of the second parenthesis in eq.  \eqref{eq:variationWardFlat} only if $\hat D^d$ contains the contribution from a level $4$ descendant which cancels that of $\pd_a \hat{T}^{ad}$ - since the bulk stress tensor is a primary.} Therefore we conclude that $\hat{A}^{ab}=0$ for $d<5$. Using this in eqs.~\eqref{eq_diffeoID} we find also $\hat{T}^{da}=0$. A similar argument rules out the existence of $\hat{B}^{\mu\nu}$ in eq.~\eqref{eq:VariationW} for $d<6$. 

Furthermore, we also notice that $\hat{T}^{ab}$ is a spin 2 operator of dimension $d-1$ and it is therefore conserved: $\pd_a \hat{T}^{ab}=0$. This means that we must have $\hat{T}^{ab}=0$ in non-trivial theories,  since otherwise we would be able to construct two set of conserved spacetime charges, signalling the presence of a decoupled sector at the boundary.\footnote{Notice that $\hat{T}^{ab}$ cannot be a descendant because $\pd_a \hat{T}^{ab}=0$ is obtained as the boundary limit of the primary operator $T^{db}$.}  

Finally,  the vanishing of $\hat{A}^{\mu\nu}$ and $\hat{T}^{\mu\nu}$ imply that the Ward identities for diffeomorphism at the boundary take the following simple form:
\begin{align}\label{eqAppNeed}
  T^{da}\vert_{x^d=0} = 0 \,,\qquad
 T^{dd}\vert_{x^d=0} =\hat  D^d  \, ,
\end{align}
where we focused again on flat space.

We end this section with three additional simple applications of the background field approach: Ward identities for internal symmetries,  the case of general defect CFTs, and the study of contact terms in correlation functions.

Consider a CFT with a continuous internal symmetry group $G$ of dimension $n_G$. In BCFTs, the boundary conditions may possibly break the symmetry to a sub-group $H\subset G$ of dimension $n_H$.  We denote with $A,B,\ldots$ the indices labelling the generators of the Lie Algebra of $G$, with $i,j,\ldots $ those of the algebra of $H$ and with $\tilde{i},\tilde{j},\ldots$ those parametrizing the coset $G/H$.
Without loss of generality, one can couple the theory to the background bulk $G$ gauge field $A_\mu^A(x)$ (with $n_G$ components) and $n_G-n_H$ boundary spurion fields $\pi^{\tilde{i}}(\sigma)$ compensating for the possible explicit symmetry breaking.  The background gauge field and spurion field transform under a gauge transformation with infinitesimal parameter $\lambda^A$ as:
\begin{align}
& A_\mu^A \to A_\mu^A + D_\mu \lambda^A \, \label{eq:InternalTrans1}\\
& \pi^{\tilde{i}} \to \pi^{\tilde{i}} +\lambda^{\tilde{i}}-\lambda^i f_{i\tilde{j}}^{\tilde{i}}\pi^{\tilde{j}}, \label{eq:InternalTrans2}
\end{align} 
where $D_\mu$ is the covariant derivative of $G$, $f^A_{BC}$ are the structure constants of the group and the spurions transform as Goldstone fields~\cite{Weinberg:1996kr}. The response of the effective action under a small variation of the sources can be parametrized as follows:
\begin{equation}
\begin{split}
\delta_G W &= \int_{\mathcal{M} } d^dx\sqrt{g}\,  \delta_G A_\mu^A
\langle J^\mu_A\rangle \\
&+ \int_{\partial\mathcal{M}} d^{d-1}\sigma \sqrt{\hat{g}}\, \left(
\delta_G A^A_\mu \langle\hat{J}^\mu_A\rangle+\delta_G \pi^{\tilde{i}} \langle \hat P_{\tilde{i}}\rangle +
n^\rho \nabla_\rho \delta_G A^A_\mu
\langle\hat{M}^\mu_A\rangle  +\cdots \right)\,.
\end{split}
\end{equation}
where $J^\mu$ is the bulk Noether current and $\hat{J}^\mu$, $\hat P$ and $\hat{M}^\mu$ encode the most general response of the boundary. We will restrict to flat space in what follows. Plugging the transformation rules \eqref{eq:InternalTrans1}, \eqref{eq:InternalTrans2},  invariance of the effective action under the gauge transformation demands (neglecting 't Hooft anomalies):
\begin{equation}\label{eq_variation_G}
\begin{aligned}
&0=- \int_{x^d>0} d^d x\,\lambda^A\partial_\mu \langle J_A^\mu\rangle \\
&+ \int_{x^d=0} d^{d-1}x \left[ \lambda^{i}
\left(-\langle J^d_i\rangle\vert_{x^d=0}-\partial_a
\langle\hat{J}^a_i\rangle\right)+ \lambda^{\tilde{i}}
\left(-\langle J^d_{\tilde{i}}\rangle\vert_{x^d=0}+\langle \hat P_{\tilde{i}}\rangle-\partial_a\langle\hat{J}^a_{\tilde{i}}\rangle\right) \right.\\
&\qquad\qquad\qquad\left.
+ \partial_d\lambda^A
\left(\langle\hat{J}^d_A\rangle-\partial_a \langle \hat{M}^a_A\rangle \right)+\partial_d^2\lambda\langle\hat{M}^d\rangle+\cdots \right]\,,
\end{aligned}
\end{equation}
where we set to zero all sources after taking the variation, since we only consider correlation functions at separated points. Gauge invariance in the bulk gives:
\begin{equation}
\pd_\mu J^\mu_A=0\,,
\end{equation}
as expected.  We also see from eq.~\eqref{eq_variation_G} that $\hat{M}^d=0$. Proceeding as before, from the unitarity bounds we conclude that $ \pd_a\hat{M}^a =0$ for $d< 5$, which leads to $\hat{J}^d = 0$, and that the boundary current $\hat{J}_A$ vanishes in a BCFT.\footnote{The boundary current for the broken generators may be the descendant of a scalar for $d>3$.  This would modify the right hand side of eq.~\eqref{eq_Ward_P},  and it would imply that every $\hat P_{\tilde{i}}$ is a linear combination of a primary and a descendant which cancels the contribution of $\partial_a\hat{J}^a_{\tilde{i}}$.  }
Therefore we find the following Ward identities:
\begin{equation}\label{eq_Ward_P}
J^d_i\vert_{x^d=0}=0\,,\qquad J^d_{\tilde{i}}\vert_{x^d=0}= \hat P_{\tilde{i}}\,.
\end{equation} 
These state that for every bulk symmetry broken by the boundary conditions there must be an operator $P_{\tilde{i}}$ of dimension $d-1$. Similarly to the displacement operator, these operators are responsible for the non-conservation of the internal charges.

All our arguments generalize almost \emph{verbatim} to generic defect CFTs. Consider a $p$-dimensional defect parametrized by coordinates $\sigma^a$ and embedding functions $X^\mu(\sigma^a)$. The response to linear perturbations of the geometry can be parametrized in terms of the bulk stress tensor $T^{\mu\nu}$ and a defect operator $\hat D^\mu$ of dimension $p+1$. The Ward identities imply:
\begin{equation}\label{eq_Ward_DCFTs}
\pd_a X^\mu \hat D_\mu=0\,,
\qquad
\nabla_\mu T^{\mu\nu}=-\delta^{d-p}(x_\bot)\hat D^\nu\,.
\end{equation}
The second eq.  in \eqref{eq_Ward_DCFTs} is equivalent to \eqref{eqAppNeed} in BCFTs, as the latter is obtained by integrating the former on a pillow geometry around the boundary.  

Finally we remark that it is also possible to analyze contact terms within this approach upon introducing sources for the operator insertions as in, e.g.,~\cite{Osborn:1993cr}. Consider for instance a $p$-dimensional linear defect in flat space and introduce a source $\hat{J}(\sigma)$ for a scalar defect operator $\hat{\mO}(\sigma)$ such that:
\begin{equation}
\frac{\delta W}{\delta \hat{J}(\sigma)}=-\frac{1}{\sqrt{\hat{g}}}\langle\hat{\mO}(\sigma)\rangle\,.
\end{equation}
Invariance of the partition functions under a combination of a diffeomorphism \eqref{eq_diffeo_var} and a boundary reparametrization \eqref{eq_repa_var} with parameters such that $\xi^\mu\vert_{\pd\mathcal{M}}=\pd_aX^\mu\zeta^a$ gives:
\begin{equation}\label{eq_WI_defect}
\left[\int d^dx \, 2\nabla_\mu \xi_\nu \frac{\delta}{\delta g_{\mu\nu}}-
\int d^p\sigma\, \zeta^a \pd_a\hat{J}\frac{\delta}{\delta \hat{J}}\right]W=0\,.
\end{equation}
Notice that with this choice of the reparametrization vectors the contribution of the displacement operator cancels. Upon taking a functional derivative with respect to the source $\hat{J}$ and setting $\xi^\mu=\delta^\mu_a$, we then find the standard Ward identity for translations parallel to the defect:
\begin{equation}
\pd_\mu\langle T^{\mu a}(x^\nu)\hat{\mO}(y^a)\ldots\rangle=
\delta^{d-p}(x_\bot)\delta^p(x^a-y^a)
\pd^a\langle\hat{\mO}(y^a)\ldots\rangle +\ldots\,,
\end{equation}
where the dots stand for contact terms associated with the other operator insertions. Recently, the generalization of the Ward identity \eqref{eq_WI_defect} to non-conformal defects played an important role in the proof of the existence of a canonically decreasing entropy function in one-dimensional defect RG flows~\cite{Cuomo:2021rkm} (see also~\cite{Affleck:1991tk,Friedan:2003yc} for a similar result in $d=2$).

\subsection{Boundary and bulk currents in weakly coupled theories}
\label{subsec:BoundaryVSBulkCurrentAndEMT}

Suppose we have a weakly coupled BCFT with action $S$ and a symmetry group $G$. As in the previous subsection, we use greek indices $\mu=1,\ldots,d$ to denote bulk indices and Latin indices $a=1,\ldots,d-1$  to denote boundary ones. Under an infinitesimal $G$-variation with spacetime dependent parameter $\varepsilon^\alpha(x)$ we must have:
\begin{equation}
\delta S=\int_{x^d>0} d^dx J^\mu _\alpha(x)\pd_\mu\varepsilon^\alpha(x)+\int_{x^d=0} d^{d-1}x
\hat{J}^a_\alpha(x)\pd_a\varepsilon^\alpha(x)\,.
\end{equation}
Standard arguments then lead to the following Ward identities:
\begin{equation}
\label{eq:ConservationRelation1}
\pd_\mu J^\mu_\alpha=0\,,\qquad
\pd_a\hat{J}^a_\alpha=-J^d_\alpha\vert_{x^d=0}\,.
\end{equation}
The above equations are consistent with having a conserved charge in the bulk theory, as expected. Note that the second relation is crucial in order to have a conserved charge associated with the bulk theory. 
In the case of translational invariance along the $x^a$  directions that are tangent to the boundary these relations read:
\begin{equation}
\label{eq:ConservationEMT1}
\pd_\mu T^{\mu a}=0\,,\qquad
\pd_a\hat{T}^{ab}=-T^{db}\vert_{x^d=0}\,.
\end{equation}
For a unitary theory, one can always improve $\hat{T}^{ab}$ to be traceless symmetric~\cite{Nakayama:2012ed}. Then the unitarity bounds imply that $T^{da}\vert_{x^d=0}=0$ and $J^d_\alpha\vert_{x^d=0}=0$ for internal symmetries. Therefore, barring the case of decoupled degrees of freedom on the boundary, in unitary CFT we only have to consider the bulk stress tensor and currents.

This raises a natural question in perturbative theories.  Indeed, it often occurs that there are bulk and  boundary currents which are separately conserved at tree-level.  In these cases, there are boundary states such that $\langle T^{\mu\nu}\rangle_{tree}=\langle J^\mu\rangle_{tree}=0$ at tree-level, whose quantum numbers are therefore measured by the boundary operators. However, we just argued above that when we consider interactions all the charges should be written in terms of the bulk operators only, apparently in contradiction with the existence of states for which these vanish at tree-level. We shall now see that this contradiction is resolved by noticing that the limit $x^d\rightarrow 0$ of the bulk current and stress tensor does not commute with the zero coupling limit.

Let us consider first the case of currents associated with internal symmetries.  We assume for simplicity that there is a unique conserved current (the generalization to many currents is straightforward). At first order in the coupling $g$, the equations of motion imply that the conservation relation \eqref{eq:ConservationRelation1} is modified at the boundary by an equation of the form:
\begin{equation}\label{eq_EOM}
\pd_a\hat{J}^a=g\hat{\mO}=-J^d\vert_{x^d=0}\, ,
\end{equation}
where for $g \to 0$ the conservation \eqref{eq:ConservationRelation1} is restored. 
Notice that the last equality is more formally written in terms of the bulk to boundary OPE as:
\begin{equation}\label{eq_bOPE}
J^d(x^d)\sim -g\hat{\mO}+\mO\left(x^{d}\right)\,.
\end{equation}
The slightly broken symmetry implies that $\hat{\mO}$ becomes a descendant of the vector $\hat{J}^a$, which acquires an anomalous dimension $\gamma_J=g^2\frac{\mathcal{N}_{\mO}}{2(d-1)}+\mO\left(g^3\right)\sim g^2$, where $\mathcal{N}_\mO$ is the normalization of the tree-level two-point function $\langle\hat{\mO}(x)\hat{\mO}(0)\rangle=\mathcal{N}_\mO/x^{2(d-1)}$~\cite{Skvortsov:2015pea,Giombi:2016hkj,DiPietro:2019hqe}.\footnote{This result for $\gamma_J$ can be derived from studying the two-point function $\langle \pd_a\hat{J}^a(x) \pd_b\hat{J}^b (0) \rangle$. }  This implies that in perturbation theory equation \eqref{eq_bOPE} gets modified according to:
\begin{equation}\label{eq_bOPE2}
J^d(x^d)\sim -(x^d)^{\gamma_J}\left[g+\mO\left(g^2\right)\right]\hat{\mO}+\mO\left(x^{d}\right)\,.
\end{equation}
Here, the $\mO(g^2)$ in parenthesis crucially refers to terms which do not depend on $x^d$. From this relation, it is already clear that the limits $x^d \to 0$ and $g \to 0$ do not commute. Notice also that we have re-summed an infinite number of (trivial) logarithms in perturbation theory to make manifest that  $J^d$ indeed vanishes  at the boundary. Finally we used the fact that $\hat{\mO}$ is the only scalar of classical dimension $(d-1)$ which can appear in the OPE, as we now prove.  Equation \eqref{eq_bOPE2} is consistent with conservation (in the limit of small $x^d$) if and only if $J^a$ contains a vector of dimension $d-2+\gamma_J$ in the boundary OPE whose divergence cancels $\pd_d J^d$. Clearly the only possibility is the current:
\begin{equation}\label{eq_bOPE3}
J^a(x^d)\sim \frac{\gamma_J+\mO\left(g^3\right)}{(x^d)^{1-\gamma_J}}\hat{J}^a+\mO\left((x^d)^0\right)\,.
\end{equation}
Notice that the current can only be used to cancel the contribution of $\hat{\mO}$, therefore no other field with the same classical scaling dimension can appear in \eqref{eq_bOPE2} unless other currents are present.  In other words, the equation of motion \eqref{eq_EOM} implies the existence of an anomalous dimension for the boundary current and therefore the OPE structure \eqref{eq_bOPE2}, \eqref{eq_bOPE3}. Similar arguments can be used to prove that only primary scalars with dimensions exactly equal to $(d-1)$ and vectors with dimension larger than $d-2$ are admissible in the bulk to boundary OPE of the current~\cite{Herzog:2017xha}.

We can finally use eq.~\eqref{eq_bOPE3} to resolve our initial puzzle. Indeed, the tree-level expectation value of the boundary current is now reproduced by integrating the bulk current over $x^d$: even in a state for which $\langle J^\mu\rangle_{tree}=0$, the OPE \eqref{eq_bOPE3} leads to the following contribution from the integration region close to the boundary
\begin{equation}\label{eq_App_int_J}
\int_0^{\mO(1)} d x^d J^a(x^d)\sim
\int_0^{\mO(1)}dx^d \frac{\gamma_J}{(x^d)^{1-\gamma_J}}\hat{J}^a+\mO\left(g^2\right)=\hat{ J}^a+\mO\left(g^2\right)\,,
\end{equation}
where in the last step we have evaluated the integral and expanded in small $g$ to find the leading order contribution. 

The same arguments can be generalized to the case of the stress tensor. In particular, assuming the equations of motion take the following form
\begin{equation}
\pd_a\hat{T}^{ab}=g \hat{V}^b=-T^{nb}\vert_{x^d=0}\,,
\end{equation}
the equations \eqref{eq_bOPE2} and \eqref{eq_bOPE3} get modified to
\begin{equation}
\begin{aligned}
&T^{nb}\sim -(x^d)^{\gamma_T} \left[g+\mO\left(g^2\right)\right] \hat{V}^b+\mO\left(x^d\right)\,,\\
&T^{ab}\sim (x^d)^{\gamma_T-1}\left[\gamma_T+\mO\left(g^3\right)\right] \hat{T}^{ab}+\mO\left((x^d)^0\right)\,,
\end{aligned}
\end{equation}
where $\gamma_T\sim g^2$ is the anomalous dimension of $\hat{T}^{ab}$.
eq.~\eqref{eq_App_int_J} generalizes to the stress tensor in a similar way.

\section{Dual boundary actions for mixed boundary conditions}\label{App_Dual_Bactions}

It is often the case that the same boundary conditions may be speficied by two different boundary actions.  When this happens, it is possible to switch from one to the other via a Legendre transform.  Here we explain this point in a simple example.

Let us consider a real scalar field in $d$-spacetime dimension, whose bulk action is:
\begin{equation}
S_{bulk}=\int_{x^d>0} d^dx\left[(\pd\phi)^2-V(\phi)\right]\,.
\end{equation}
We consider boundary conditions at $x^d=0$ of the following form
\begin{equation}\label{eq_App_Bdry_condition0}
f(\phi,\pd_d\phi)\vert_{x^d=0}=0\,,
\end{equation}
where $f$ is an arbitrary function of $\phi$ and its normal derivative $\pd_d\phi$ at the boundary. eq.~\eqref{eq_App_Bdry_condition0} can be equivalently solved for $\phi$ or for $\pd_d\phi$:
\begin{equation}\label{eq_App_Bcondition1}
\phi\vert_{x^d=0}=P(\pd_d\phi)\vert_{x^d=0}\qquad\iff\qquad
\pd_d\phi\vert_{x^d=0}=G(\phi)\vert_{x^d=0}\,.
\end{equation}
Correspondingly, we can specify the boundary condition \eqref{eq_App_Bdry_condition0} via two different but equivalent boundary actions. The simplest option is to just consider:
\begin{equation}\label{eq_App_Bdry_action_1}
S_{bdry}=-\int_{x^d=0} d^{d-1}x \,W(\phi)\,,\qquad
W'=G\,.
\end{equation}
It is then easy to check that the variation of $S+S_{bulk}$ imposes precisely the boundary condition \eqref{eq_App_Bcondition1} written in the form $\pd_d\phi=G(\phi)$.

We may dualize the boundary action \eqref{eq_App_Bdry_action_1} integrating-in an auxiliary field $\lambda$ as follows:
\begin{equation}
S_{bdry}=-\int_{x^d=0}d^{d-1}x\left[W(\lambda)+\pd_d\phi(\phi-\lambda)\right]\,.
\end{equation}
To check the equivalence with eq.~\eqref{eq_App_Bdry_action_1} it is useful to notice that the variation of $\phi\pd_d\phi$ produces a term $\delta\phi\pd_d\phi$ which cancels the boundary term from the variation of the bulk action. 
The variation $\delta(\pd_d\phi)$ at the boundary then sets $\lambda=\phi$. 
Integrating out $\lambda$ instead we obtain the following action
\begin{equation}
S_{bdry}=-\int_{x^d=0} d^{d-1}x\left[\phi\pd_d\phi+\widetilde{W}\left(\pd_d\phi\right)\right]\,,
\end{equation}
where $\widetilde{W}$ is the Legendre transform of $W$:
\begin{equation}
\widetilde{W}(\pd_d\phi)=W(\lambda^*)-\lambda^*\pd_d\phi
\quad\text{with}\quad
G(\lambda^*)=\pd_d\phi\,.
\end{equation}
It is simple to check that the variation $\delta(\pd_d\phi)$ imposes eq.~\eqref{eq_App_Bcondition1} in the form $\phi=P(\pd_d\phi)$.

As a simple illustration, we can consider linear boundary conditions
\begin{equation}
\left(\pd_d\phi-\mu\phi\right)_{\pd\mathcal{M}}=0\,.
\end{equation}
Following the previous steps we find that these can be represented in two equivalent ways
\begin{equation}
S_{bdry}^{(1)}=-\int d^{d-1}x\frac{\mu}{2}\phi^2\qquad
\text{or}\qquad
S_{bdry}^{(2)}=-\int d^{d-1}x\left[\phi\pd_d\phi-\frac{(\pd_d\phi)^2}{2\mu}\right]\,.
\end{equation}
We notice that the $S^{(1)}_{bdry}$ is apparently singular in the limit $\mu\rightarrow\infty$, corresponding to Dirichlet boundary conditions, but it is trivial for $\mu=0$ corresponding to Neumann boundary conditions. Conversely, $S^{(2)}_{bdry}$ is singular for $\mu\rightarrow 0$ but it is regular for Dirichlet boundary conditions $\mu\rightarrow\infty$.

\section{The conformal superfluid action with a boundary}\label{AppBdryAction}

In this section we discuss how to construct the EFT action \eqref{eq_action_LO} on $\mathds{R}\times\mathds{H}S^{d-1}$. Let us first recall how to obtain the bulk action \eqref{eq_action_LO_bulk}. To this aim it is enough to notice that the most general $U(1)$, diffeomorphism and Weyl-invariant action for a scalar is obtained contracting $\pd_\mu\chi$ and geometric invariants obtained out of a rescaled metric $\tilde{g}_{\mu\nu}=(\pd\chi)^2g_{\mu\nu}$. We distinguish these from the one constructed out of the standard metric with a tilde. Discarding terms which vanish on the leading order equations of motion, the leading terms read:
\begin{equation}
S_{bulk}=\int_{\mathcal{M}} d^dx\sqrt{\tilde{g}}\left[c_1+c_2\tilde{R}+c_3\tilde{R}^{\mu\nu}\pd_\mu\chi\pd_\nu\chi+\mO\left(\tilde{\nabla}^4\right)\right]\,,
\end{equation}
whose expansion coincides with eq.~\eqref{eq_action_LO_bulk}. Notice that we are also assuming that the bulk theory is parity invariant, otherwise in $d=3$ it would be possible to write terms which are first order in derivatives in terms of the gauge field dual to $\chi$~\cite{Cuomo:2021qws}.

We now discuss the boundary conditions.  Classically, the boundary conditions should provide enough information to solve a second order differential boundary value problem.  At a quantum-level, we can think of a boundary as specifying a state in the appropriate quantization~\cite{Cardy:2004hm}. Either way, since $\chi$ is a shift invariant scalar, the boundary condition can only specify the value of its normal derivative.  To do so in the most general way compatible with the $d-1$-dimensional conformal group, we can parametrize the boundary conditions with a boundary action.  Since the latter lives in $(d-1)$ dimension, it is necessarily $\mO(1/\mu)$ suppressed with respect to the bulk action. Therefore, to leading order in the derivative expansion, the boundary condition arises from the variation of the bulk action and reads:
\begin{equation}\label{eq_App_Bdry_EOM}
n^\mu j_\mu\vert_{\pd\mathcal{M}}=n^\mu(\pd\chi)\pd_\mu\chi\vert_{\pd\mathcal{M}}\simeq 0\,.
\end{equation}
We can use this condition to discard all terms proportional to $n^\mu\pd_\mu\chi$ at the boundary.\footnote{This is because, in perturbation theory, their effect is the same as that of higher derivative operators which do not involve normal derivatives.} This implies that to leading order in derivatives we can only write one term compatible with Weyl invariance:
\begin{equation}\label{eq_App_Bdry_Baction}
S_{bdry}=\int d^{d-1}x\sqrt{\hat{g}}\,b_1(\hat{\pd}\chi)^{d-1}+\ldots\,.
\end{equation}
We will now show that this action does not receive correction to first subleading order in derivatives.

To construct higher order terms it is convenient to work in terms of the rescaled metric $\tilde{g}_{\mu\nu}$ as before.
We use the same notation of appendix \ref{subsec:BackgroundFieldsApproach} for geometric quantities, including a tilde to denote when they are constructed from the rescaled metric, for instance $\tilde{n}_\mu=(\pd\chi)n_\mu$, $\hat{\tilde{g}}_{ab}=\hat{g}_{ab}(\pd\chi)^2$, etc. To first order in derivatives there are three Weyl-invariant operators:
\begin{equation}
\hat{\mO}_1=\hat{\tilde{\nabla}}^a\pd_a\chi\,,\qquad
\hat{\mO}_2= \pd_a\chi\pd_b\chi\tilde{K}^{ab}\,,\qquad
\hat{\mO}_3=\tilde{K}_{ab}\hat{\tilde{g}}^{ab}\,,
\end{equation}
where $\tilde{K}_{ab}=-\tilde{n}_\mu\hat{\tilde{\nabla}}_a\pd_b X^\mu$ is the rescaled extrinsic curvature.  The first operator is clearly a total derivative. Using that the extrinsic curvature of the equator vanishes, we find that
\begin{equation}
\tilde{K}_{ab}=\hat{g}_{ab}\frac{n^\mu\pd_\mu(\pd\chi)}{(\pd\chi)^2}\,,
\end{equation}
and therefore all terms proportional to the extrinsic curvature vanish on the leading order boundary condition \eqref{eq_App_Bdry_EOM}.

Finally in odd spacetime dimensions we need to consider the Wess-Zumino term to reproduce the boundary Weyl anomaly in the EFT.  For instance in $d=3$ this depends on two coefficients $b_a$ and $d_a$~\cite{Jensen:2015swa}.  While the Wess-Zumino term is usually written for a background dilaton field, here its role is played by the dynamical composite field $\log (\pd\chi)$. Since $\chi$ itself may contribute to the Weyl anomaly, it's the difference of central charges between the UV BCFT and the superfluid theory that shows up in the following expressions:\footnote{This phenomenon in the context of 2d bulk CFT was demonstrated through explicit computation in~\cite{Komargodski:2021zzy}.}
\begin{equation}
\begin{split}
S_{WZ}=&-\frac{1}{24\pi}\int_{\pd \mathcal{M}} d^2x\sqrt{\hat{g}}\log (\pd\chi)\left[\Delta b_a 
\hat{\mathcal{R}}+\Delta d_a \left(K_{ab}-\text{trace}\right)^2\right]\\
&+\frac{\Delta b_a}{24\pi}\int_{\pd \mathcal{M}} d^2x\sqrt{\hat{g}}\,\pd_a\log(\pd\chi)\hat{g}^{ab}
\pd_b\log(\pd\chi)
\,.
\end{split}
\end{equation}
Notice that the first line vanishes identically on the boundary of $\mathcal{M}=\mathds{R}\times\mathds{H}S^{d-1}$.
In general the Wess-Zumino term is always of order $\mO(\mu^0)$ in the EFT and is therefore suppressed by $(d-1)$ derivatives with respect to the term in eq.~\eqref{eq_App_Bdry_Baction}. Therefore we conclude that all additional higher derivative terms in the boundary action \eqref{eq_App_Bdry_Baction} are suppressed by two or more derivatives.  It would be interesting to understand, if the anomaly terms lead to distinctive features in the EFT despite this high suppression.

\section{Details on the epsilon expansion}\label{App_D}

\subsection{Anomalous dimensions at small \texorpdfstring{$\lambda_* Q$}{lambda Q}}\label{appendix_diagrams}

In this subsection we explain how to obtain the results \eqref{eq_Neumann_anomalous} and \eqref{eq_anomalous_Dir} for the one-loop anomalous dimensions of the lowest dimensional charge $Q$ boundary operators in the Wilson-Fisher fixed point with Neumann and Dirichlet boundary conditions. As a reminder, one defines the wave-function renormalization $\hat{Z}_Q$ of a bare operator $\hat{\mO}_Q$~\cite{Kleinert} isolating the divergent part of its correlation functions with other (renormalized) operators:
\begin{equation}
\langle \hat{\mO}_Q(x)\ldots\rangle=\hat{Z}_Q\underbrace{\langle[\hat{\mO}_Q](x)\ldots\rangle}_{=\text{finite}}\,,\qquad
x^d=0\,,
\end{equation}
where $\hat{\mO}_Q=\hat{\phi}^Q$ for the Neumann boundary conditions, while $\hat{\mO}_Q=(\pd_d\hat{\phi})^Q$ for Dirichlet. The anomalous dimension then follows from:
\begin{equation}\label{eq_formula_gamma}
\hat{\gamma}_Q=\frac{\pd \log \hat{Z}_Q}{\pd\lambda}\beta_\lambda\,,
\end{equation}
where the $d$-dimensional beta function is given in eq.~\eqref{eq_beta_function}.

\begin{figure}[t]
   \centering
		\subcaptionbox{  \label{fig:diagram1BCFT}}
		{\includegraphics[width=0.35\textwidth]{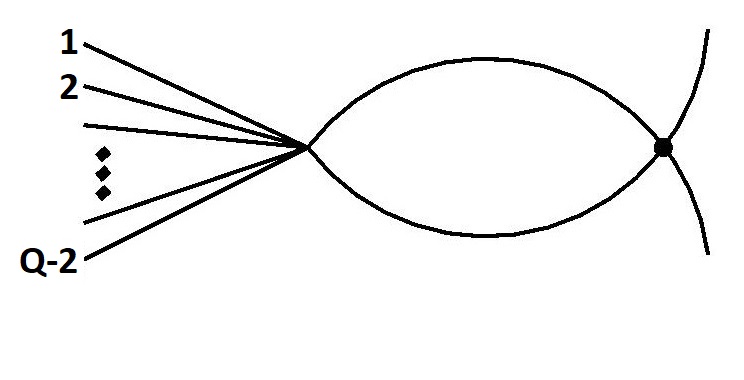}}
	\qquad \qquad \qquad \qquad 
		\subcaptionbox{ \label{fig:diagram2BCFT}}
		{\includegraphics[width=0.35\textwidth]{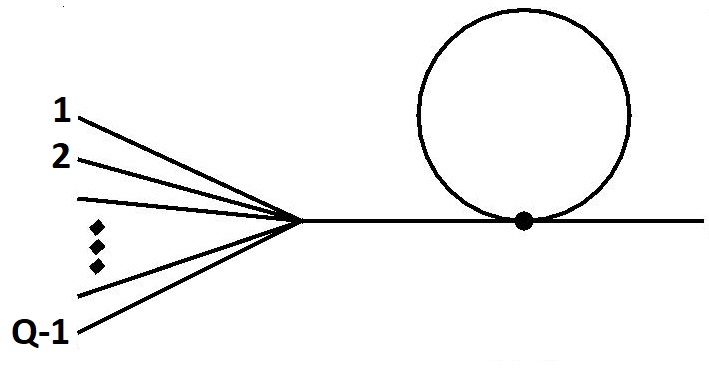}}
        \caption{ Feynman diagrams that contribute one-loop corrections to the anomalous dimension \eqref{eq_Neumann_anomalous}.    }
\label{fig:DiagramsBetaFunctionO2Model}
\end{figure}

Let us consider first Neumann boundary conditions $\pd_d\phi\vert_{x^d=0}=0$.  The $\phi$ propagator in this case is~\cite{McAvity:1995zd}:
\begin{equation}
\label{eq:NeumannProp}
G_N(y,x)=\frac{1}{(d-2)\Omega_{d-1}(y-z)^{d-2}}+\frac{1}{(d-2)\Omega_{d-1}(y-\bar{z})^{d-2}}\,,\qquad
\bar{z}=(z^1,\ldots,z^{d-1},-z^d)\,.
\end{equation}
To one-loop accuracy we consider one insertion of the interaction term in a generic correlator containing the operator $\hat{\phi}^Q$:
\begin{equation}
\begin{split}
\langle \hat{\phi}^Q(x)\ldots\rangle &\simeq\langle \hat{\phi}^Q(x)\ldots\rangle_{free}-
\frac{\lambda}{4}\langle\hat{\phi}^Q(x)\int_{y^d\geq 0} d^dy(\bar{\phi}\phi)^2(y)
\ldots\rangle_{free} \,,
\end{split}
\end{equation}
where the subscript stresses that all matrix elements are evaluated via Wick contraction using eq.~\eqref{eq:NeumannProp}.
To extract the divergent contribution, we consider all possible contractions of the interaction term with the operator $\hat{\phi}^Q(x)$. To this order we obtain:
\begin{equation}
\begin{split}
\langle \hat{\phi}^Q(x)\ldots\rangle & \simeq\langle \hat{\phi}^Q(x)\ldots\rangle_{free}-
\lambda Q\langle\hat{\phi}^{Q-1}(x)\int d^dy\, G_N(x,y)G_N(y,y)\phi(y)
\ldots\rangle_{free} \\[6pt]
&-
\frac{\lambda Q(Q-1)}{4}\langle\hat{\phi}^{Q-2}(x)\int_{y^d\geq 0} d^dy\, G_N^2(x,y)\phi^2(y)
\ldots\rangle_{free}+
\text{finite}\,,
\end{split}
\end{equation}
These contractions correspond to the diagrams in figure \ref{fig:DiagramsBetaFunctionO2Model}.
Logarithmic divergences\footnote{There is a power divergence for $y^d\rightarrow 0$ with $y_{||}$=fixed, but its contribution vanishes in dimensional regularization.} arise only from the limit $y^d\rightarrow 0$ and $y_{||}\rightarrow x_{||}$.  Therefore we can extract them by expanding $\phi(y)=\phi(x)+\ldots$ and introducing an arbitrary IR cutoff in the integration. This implies that we can absorb the divergences in the correlator with the following wave-function renormalization:\footnote{To one-loop accuracy the anomalous dimension are scheme-independent even away from the fixed point.}
\begin{equation}
\begin{split}
\hat{Z}_Q&=1-\lambda Q\int' d^dy\,G_N(x,y)G_N(y,y)-\frac{\lambda Q(Q-1)}{4}\int' d^dy\,G_N^2(x,y)\,,\\
&=1-\frac{\lambda}{16\pi^2 \varepsilon}\left(Q^2-2 Q\right)+\text{finite}\,,
\end{split}
\end{equation}
where the prime on the integral serves as a reminder that an IR cutoff of our choice must be introduced in the integration; this determines the finite scheme-dependent terms in the second line.  Notice also that, in dimensional regularization, the propagator $G_N(y,y)$ at coincident points is obtained simply discarding the first divergent term in eq.~\eqref{eq:NeumannProp}.
Using eq.~\eqref{eq_formula_gamma} with $\beta_{\lambda}\simeq -\varepsilon\lambda$ we find the result \eqref{eq_Neumann_anomalous}.  

The same procedure can be used to extract the wave-function of the operator $(\pd_d\hat{\phi})^Q$ for Dirichlet boundary conditions. In this case the propagator is given by
\begin{equation}
\label{eq:propDirichlet}
G_D(y,x)=\frac{1}{(d-2)\Omega_{d-1}(y-z)^{d-2}}-\frac{1}{(d-2)\Omega_{d-1}(y-\bar{z})^{d-2}}\,,
\end{equation}
where $\bar{z}$ is defined as in eq.~\eqref{eq:NeumannProp}.  Proceeding as before we then find
\begin{equation}\label{eq_Z_Dirichlet}
\begin{split}
\hat{Z}_Q&=1-\lambda Q\int' d^dy\, y^d \partial_d G_D(x,y)G_D(y,y)-\frac{\lambda Q(Q-1)}{4}\int' d^dy\, y^dy^d\left(\partial_d G_D(x,y)\right)^2\,,\\
&=1-\frac{\lambda}{32\pi^2\varepsilon}\left(Q^2-3 Q\right)+\text{finite}\,,
\end{split}
\end{equation}
where again we introduced an arbitrary IR cutoff in the integration and neglected the scheme-dependent contributions in the second line.  Notice that, while the propagator \eqref{eq:propDirichlet} vanishes when $x$ is at the boundary, its normal derivative does not. Using eq.~\eqref{eq_formula_gamma} we then obtain the result \eqref{eq_anomalous_Dir} in the main text.

As a consistency check, we note that the results \eqref{eq_Neumann_anomalous} and \eqref{eq_anomalous_Dir} are in agreement with those obtained in~\cite{McAvity:1995zd} for $Q=1$.

\subsection{Casimir energy for large \texorpdfstring{$\lambda_* Q$}{lambda n} in the epsilon expansion}\label{App_Casimir_epsilon}

We would like to evaluate the sum we encountered in \eqref{eqDelta0neumann} for large $\mu R$, which we repeat here:
\es{sigellSum}{
&S(\mu R)\equiv \frac12\sum_{\ell=1}^\infty \sigma(\ell)\,,\\
&\sigma(\ell)=\frac{(\ell+1) (\ell+2)}{2}R\left[\omega_+(\ell)+\omega_-(\ell)\right]
-\ell^3-4 \ell^2-\ell \left(\mu^2 R^2+4\right)-2 \mu^2 R^2+\frac{5 \left(\mu^2 R^2-1\right)^2}{8 \ell}\,,
}
where we omitted the $*$ notation of the main text to avoid clutter. The large $\mu R$ limit is a flat space limit, terms with $\ell\sim \mu R$ should dominate and the result should be expressible by integrals instead of sums. We realize these expectations below.

First we want to perform the sum for ``low $\ell$'s'', from $\ell=1$ to $\ell=a\mu R$, where $a\ll 1$ such that $a\mu R$ is a large integer. We expand the summand for large $\mu R$, to get:
\es{lowellExp}{
\sigma(\ell)=&\le(\mu R\ri)^4\le[{\frac{5}{8}\ell}+{4\ell^2+8\ell+\frac{5}{\ell}}\,{\frac{1}{\le(\mu R\ri)^2}}+\sqrt{\frac{3}{2}}\,(\ell+1)(\ell+2)\,{\frac{1}{\le(\mu R\ri)^3}}+\dots\ri]\,.
}
The terms can be summed in closed form and expanded for small $a$. 
The final result takes the form
\es{lowellExp2}{
S_\text{low}(\mu R)=& \frac12\sum_{\ell=1}^{a\mu R} \sigma(\ell)\\
=&\le(\mu R\ri)^4\le[{\frac{5}{16}}\log(a\mu R \,e^{\ga_E})-\frac14 a^2+\dots\ri.\\
&+{\frac{1}{\mu R}}\le({\frac{5}{32 a}}-{\frac{5}{4}}a+\dots\ri)\\
&+{\frac{1}{ \le(\mu R\ri)^2}}\le(-{\frac{5}{ 192 a^2}}-{\frac{5}{ 8}}\log(a\mu R \,e^{\ga_E})+\dots\ri)\\
&+{\frac{1}{ \le(\mu R\ri)^3}}\le(-{\frac{5}{ 8 a}}+\le({\frac{47}{ 24\sqrt{3}}}-\frac83\ri)a+\dots\ri)+\dots \Big ]\,,
}
where we hoped to convey the structure of this double expansion and demonstrate that to obtain the terms in \eqref{eq_epsilon_Delta0_large} we only need to expand to $O(\mu R)$ in \eqref{lowellExp}.

Next, we want evaluate the sum for ``high $\ell$'s'', from $\ell=a\mu R+1$ to $\ell=\infty$. Note that $S_\text{high}(\mu R)$ is supposed to have such an $a$-dependence that it completely cancels the $a$-dependence of $S_\text{low}(\mu R)$. We proceed by introducing a variable to be regarded as (dimensionless) flat space momentum, $k\equiv \ell/(\mu R)$, and write the Euler-Maclaurin formula as
\es{highell}{
S_\text{high}(\mu R)&= \frac12\sum_{a\mu R+1}^\infty \sigma(\ell)\\
&={\frac{\mu R}{ 2}}\int_{a}^\infty dk \ \Sigma(k)-{\frac{\Sigma(a)}{ 4}}-\sum_{m=1}{\frac{B_{2m}}{ 2 (2m)! (\mu R)^{2m-1}}}\, \Sigma^{(2m-1)}(a)\,,\\
\Sigma(k)&\equiv \sig(k\mu R)\,.
}
The function $\Sigma(k)$ is an explicit function of $\mu R$, and can be straightforwardly expanded as 
\es{SigExp}{
\Sigma(k)&={\frac{1}{ \mu R}}\, \Sigma_1(k)+{\frac{1}{ (\mu R)^2}}\, \Sigma_2(k)+\dots\,,\\
\Sigma_1(k)&=-k^3-k+{\frac{5}{ 8 k}}+\frac12 k^2\le(\sqrt{k^2+3+\sqrt{4k^2+9}}+\sqrt{k^2+3-\sqrt{4k^2+9}}\ri)\,,
}
where we have determined the expansion up to $\Sigma_4(k)$, but will spare the reader from the explicit expressions. 

All we would have to do is to plug this expression back into \eqref{highell} and expand for small $a$. This would be rather cumbersome because of the ${5/ 8 k}$ term in \eqref{SigExp} that would make the this a singular expansion near $k=0$. Subtracting this term is not an option either, since this would ruin the fast decay of $\Sigma_1(k)$ for $k\to \infty$. Instead, we look for a subtraction that while making the integrand regular as $k\to0$ does not spoil the $k\to \infty$ asymptotics and that is simple enough so that it can be exactly summed over $\ell$.  The following subtraction fits the bill:
\es{SigHat}{
\widehat{\Sigma}_1(k)&=\Sigma_1(k)-{\frac{5}{8k\le(1+8k^2/5\ri)}}\,,
}
where the ${8/ 5}$ prefactor of $k^2$ is inessential, but makes some subsequent expressions simpler.\footnote{At higher orders in $1/(\mu R)$ we can use the same subtraction with its overall coefficient adjusted. In particular, $\Sigma_{2,4}$ does not require subtraction and $\widehat{\Sigma}_3(k)=\Sigma_3(k)+{\frac{5}{4k\le(1+8k^2/5\ri)}}$.} Now, we can plug $\widehat{\Sigma}_i(k)$ into \eqref{highell}, and preform the small $a$ expansion. We find that the $a$ dependent terms drop out from the sum of $S_\text{low}(\mu R)+S_\text{high}(\mu R)$.\footnote{In principle, it would be sufficient to perform the small $a$ expansion only until we reach a positive power of $a$ at a given order in $1/(\mu R)$, since by construction the result cannot depend on $a$ and $a\ll 1$. However, in practice we found it a very useful check to observe the cancellation of higher powers of $a$ between the low and high $\ell$ sums.}  The final result reads
\es{SFinal}{
S(\mu R)=&\le(\mu R\ri)^4\le[{\frac{5}{16}}\log(\mu R )+{\frac{5}{32}}\le(2\ga_E-\log\le(\frac{8}{5}\ri)\ri)+\frac12\int_{0}^\infty dk \ \widehat{\Sigma}_1(k)\ri]\\
&+\le(\mu R\ri)^3 \frac12\int_{0}^\infty dk \ {\Sigma}_2(k)\\
&+\le(\mu R\ri)^2\le[-{\frac{5}{8}}\log(\mu R )-{\frac{5}{16}}\le(2\ga_E-\log\le(\frac{8}{5}\ri)\ri)+{\frac{13}{ 24}}+\frac12\int_{0}^\infty dk \ \widehat{\Sigma}_3(k)\ri]\\
&+\le(\mu R\ri)\le[{\frac{5\sqrt{3}}{ 8 \sqrt{2}}}+\frac12\int_{0}^\infty dk \ {\Sigma}_4(k)\ri]\\
&+\dots
}
The above procedure is another application of matched asymptotic series expansion that we used in this paper repeatedly.

Next we plug in $d=4$ into \eqref{MuLamRel} to relate $\mu R$ to $\lam n$:
\es{MuLamRel2}{
(\mu R)^3-{\mu R }=\frac{\lambda Q}{4\pi^2 }\,,\quad\implies\quad
\mu R =\le(\frac{\lambda Q}{4\pi^2 }\ri)^{1/3}+\frac13\, \le(\frac{\lambda Q}{4\pi^2 }\ri)^{-1/3}+\dots\,.
}
Plugging the second line into \eqref{eqDelta0neumann} and \eqref{SFinal}, we finally obtain \eqref{eq_epsilon_Delta0_large}.

\bibliography{Biblio}
	\bibliographystyle{JHEP.bst}

\end{document}